\documentclass[journal]{IEEEtran}

\usepackage{mathtools,amssymb,lipsum, nccmath}

\usepackage{cuted}
\usepackage{cite}
\usepackage{rotating}
\usepackage{tabularx}
\usepackage{amsmath,amssymb,amsfonts}
\usepackage{mathtools}
\usepackage[linesnumbered,ruled,vlined]{algorithm2e}
\usepackage{algcompatible}

\SetCommentSty{mycommfont}
\usepackage[flushleft]{threeparttable}
\usepackage{footnote}
\usepackage{nicematrix,enumitem,booktabs}
\makesavenoteenv{tabular}
\usepackage{xcolor}
\usepackage{color}
\usepackage{url}
\usepackage{mdwmath}
\DeclareUnicodeCharacter{202F}{ }
\usepackage[all]{xy}
\usepackage{array}
\usepackage[caption=false,font=footnotesize]{subfig}
\providecommand{\subfloat}[2][]{#2}
\usepackage{stfloats}
\usepackage{bbm} %Indicator function
\usepackage{adjustbox}
\usepackage{orcidlink}
\usepackage{graphicx}
\graphicspath{{./figures/}}
\DeclareGraphicsExtensions{.pdf,.jpeg,.eps}
\usepackage{multicol}
\usepackage{graphicx}
\usepackage{setspace}

\SetKwInput{KwData}{Input}
\SetKwInput{KwResult}{Output}

\begin{document}
	\title{Leveraging Space-Time Synchronization for Ultra-Spot Detection in mmWave/THz UAV-to-UAV Communications}
	
	\author{Phuc Duc~Nguyen\,\orcidlink{0000-0002-7136-8924}~\IEEEmembership{Senior Member,~IEEE,}
		Ryosuke~Isogai\,\orcidlink{0009-0001-3743-4262}~\IEEEmembership{Member,~IEEE,} Keitarou Kondou\orcidlink{0009-0006-2762-8895}, Satoshi Yasuda\,\orcidlink{0000-0002-0196-6734}, Nobuyasu Shiga\,\orcidlink{0000-0002-6349-7662}~\IEEEmembership{Member,~IEEE,}  
	and~Yozo~Shoji\,\orcidlink{0000-0003-3523-9620}~\IEEEmembership{Member,~IEEE}% <-this % stops a space
		\thanks{The paper has been accepted and published by IEEE Transactions on Vehicular Technology. (Corresponding author: Phuc Duc Nguyen, phucdnguyen@nict.go.jp)}
		\thanks{P. D. Nguyen, K. Kondou, Y. Shoji are with Social-ICT System Laboratory \{phucdnguyen,keitarou.kondou,shoji\}@nict.go.jp, (R. Isogai) ryosuke.isogai@seiko-sfc.co.jp; S. Yasuda, N. Shiga are with Space-Time Standards Laboratory \{sayasuda, shiga\}@nict.go.jp, National Institute of Information and Communications Technology (NICT), 4-2-1, Nukui-Kitamachi, Koganei 184-8795, Tokyo, Japan.}% <-this % stops a space
        \thanks{A portion of this work was presented at the 2024 International Symposium on Personal, Indoor and Mobile Radio Communications (PIMRC'24), and IEEE 13th Global Conference on Consumer Electronics (GCCE 2024). This journal version extends the previous works by providing theoretical proofs of the spatial diversity effects, offering a detailed analysis of the proposed algorithms, and including more comprehensive simulation and experimental results.}
	}

    % \markboth{IEEE Transactions on Vehicular Technology, Vol. xx, No. x, x 2026}%
	% {Shell \MakeLowercase{\textit{et al.}}: Bare Demo of IEEEtran.cls for IEEE Journals}
	
	\maketitle
	
	\begin{abstract}
    In UAV-to-UAV communication, airborne UAVs need to detect the location and direction of ultra-high-speed millimeter-wave (mmWave) and Terahertz (THz) coverage areas, referred to as ultra-spots. This predictive capability allows UAVs to optimally adjust their flight paths, altitude, and velocity, thereby maximizing the utilization of ultra-spot services. A space-time synchronization technique employing multiple Wireless Two-way Interferometry devices (multi-Wi-Wi) is proposed in this paper to detect mmWave/THz ultra-spot locations during UAV operations. This paper proposes an algorithm that estimates the likelihood of nearby ultra-spots by considering the UAV flight route and ultra-spot direction, and by sharing location and pose information among UAVs in the network via a 920\,MHz wireless communication link. For the first time, \textcolor{black}{this work} addresses the problem of optimizing UAV flight routes to maximize ultra-spot utilization. To address the inherent challenges of Wi-Wi, such as phase data unreliability, RSSI attenuation, or packet loss caused by obstructions from the UAV’s own body, this study proposes the use of multiple Wi-Wi devices equipped with antennas positioned at different positions around the arms of the UAV to leverage spatial diversity effects. The proposed method’s effectiveness is confirmed through experimental data derived from real-world UAV-to-UAV communication tests. An error \textcolor{black}{of 37.16\,cm was observed experimentally in ultra-spot location estimation, corresponding to 186\,ms error in temporal prediction} of ultra-spot entry from an in-flight UAV, demonstrating its effectiveness in addressing ultra-spot detection challenges in mmWave communication. 
	\end{abstract}
	
	
	\IEEEpeerreviewmaketitle
	
\section{Introduction}
\label{sec:introduction}
\IEEEPARstart{H}{i}gh-speed communication systems utilizing mmWave and Terahertz (THz) waves enable the transmission of large volumes of data in secure environments to designated devices; however, they suffer significant power attenuation at ultra-high frequencies compared to microwave-based technologies. To address this, highly directional and energy-focused narrow beams need to be optimally formed and aligned to a target communication device using beamforming techniques \cite{xing2021millimeter,tripathi2021millimeter}. To enable high-speed communication, a pair of ultra-narrow beams, ranging from a few centimeters to over a meter in width, must be precisely aligned in position and angle between the transmitter and receiver. This alignment forms ultra-spots, spatially confined zones that are flexible in time and space, support high-capacity, low-latency, and ultra-secure transmissions, and are reserved for pre-authorized users \cite{jiang2024terahertz}. In UAV-to-UAV communication, early detection of these ultra-spots while the UAVs are in flight is crucial for path planning\cite{basil2025performance,zhai2025energy}, and beam steering before the UAVs approach the ultra-spot\cite{khemiri2025robust,kim2025two}. This enables the activation of mmWave/THz transceivers at a precise time, thereby conserving energy and enhancing communication efficiency. Especially in scenarios where two UAVs communicate at high speed while passing each other, gigabytes of data are transmitted within a very brief timeframe, typically lasting only a few seconds or less than one second \cite{isogai2023challenges, nguyen2025prediction}. 
Additionally, one of the primary challenges for UAVs or high-speed moving objects, such as cars or trains, is maintaining continuous, uninterrupted high-speed communication through beam tracking \cite{ding2021context, huang20203d, ke2019position}. The inherently narrow beamwidth of mmWave/THz signals, combined with the mobility of devices such as vehicles or UAVs, makes beam tracking and alignment highly challenging, as any deviation in position or beam angle estimation can lead to communication failure. Therefore, there is a pressing need for a technique that can not only achieve high-precision localization but also synchronize beam steering timing between the transmitter and receiver to address the above issue.
 \begin{figure}[!t]
	\centerline{\includegraphics[width=\columnwidth]{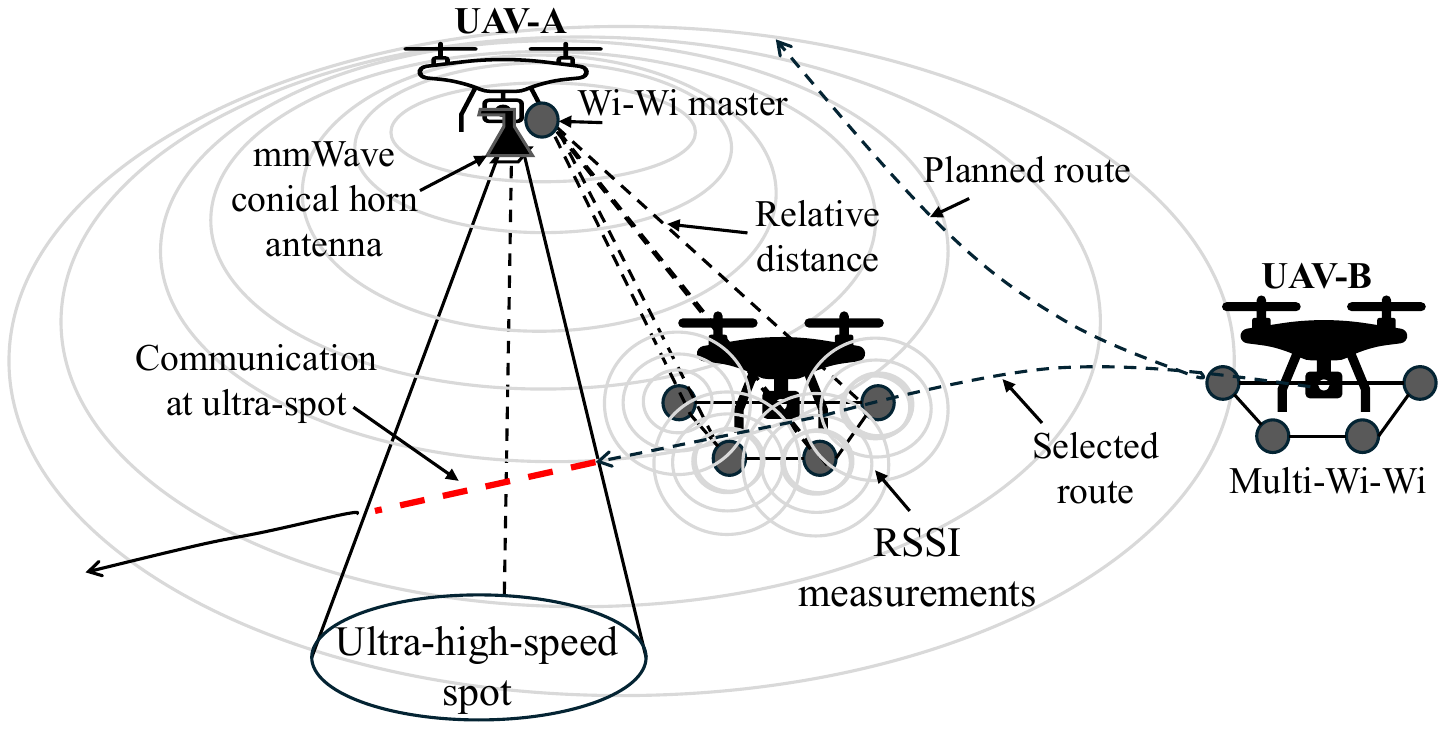}}
	\caption{The concept involves using RSSI measurements and continuous phase variations from four Wi-Wi wireless devices mounted on the UAV-B to determine the location and angle of the ultra-spot. This enables early detection of the deviation in UAV-B’s orientation from the direction of the ultra-spot created by UAV-A, allowing UAV-B to adjust its flight path to approach the ultra-spot.}
	\label{fig:01}
\end{figure}

%already revised
The primary challenge in ultra-spot detection lies in accurately identifying the target ultra-spot location while the UAV is in flight. Although the Global Navigation Satellite System (GNSS) is widely employed for positioning \cite{li2015accuracy}, its inherent error margin, which can extend to several meters, makes it unsuitable for detecting mmWave/THz ultra-spots due to their ultra-narrow size. While Real-Time Kinematic (RTK) technology can enhance GNSS positioning accuracy to the centimeter level, it relies on base stations, which incur additional costs \cite{medina2018kalman}. Furthermore, current positioning technologies are limited in indoor environments. Therefore, there is a pressing need for a positioning method that can seamlessly transition between indoor and outdoor environments while maintaining high accuracy for ultra-spot detection. This requirement is particularly critical in scenarios such as detecting ultra-spots in tunnels or under extreme conditions where GNSS and RTK-GNSS are unavailable.

%already revised
To address the challenges above, we propose an ultra-spot detection approach based on space-time synchronization, comprising two main components: spatial synchronization and temporal synchronization. This approach integrates several technologies, with Wireless Two-way Interferometry (Wi-Wi) being the primary one. Wi-Wi not only enables precise time synchronization but also provides spatial synchronization by frequently detecting changes in distance between master and slave Wi-Wi devices. Specifically, we share the location and velocity information of UAVs within the network over a dedicated 920\,MHz wireless channel before flight, and continuously update their inter-UAV distances during flight using Wi-Wi technology \cite{shiga2017demonstration,yamasaki2021delay,akasaka2023implementation}. This technology measures the carrier phase difference caused by signal propagation delays. For precise time synchronization, the principle of two-way time and frequency transfer (TWTFT) measures the clock offset and propagation delay between two distant clocks. Synchronization is achieved by exchanging time signals in both directions and comparing the internal clock time with the received signal at each end, as reported by Yasuda et al. \cite{yasuda2019horizontal}. The Wi-Wi system operates at a carrier frequency of approximately 920\,MHz with a channel bandwidth of 200\,kHz, providing a phase output data rate of 20\,Hz. Moreover, the Wi-Wi devices are IEEE 802.15.4g compliant, making them cost-effective and potentially as affordable as standard commercial wireless communication devices.

In addition to time synchronization ability with precision down to 10\,ns, Wi-Wi can enable space synchronization, or more simply explained, its ability to measure distance with an accuracy of less than 1\,cm \cite{isogai2023extremely}. However, Wi-Wi technology has limitations, such as Received Signal Strength Indicator (RSSI) degradation and high packet loss rates in environments with obstacles, particularly those posed by the UAV itself. These obstacles cause instability in Wi-Wi's phase variation values, leading to inaccurate distance variation calculations. In this paper, we propose a mechanism to address this drawback by utilizing data from multi-Wi-Wi placed at different locations on the UAV, thereby enhancing the robustness of the ultra-spot detection method.

\textcolor{black}{This paper introduces several novel contributions that collectively advance the emerging field of UAV‑assisted ultra‑spot sensing. First, we present the multi‑Wi‑Wi concept, an innovative configuration that deploys multiple Wi‑Wi devices on UAV platforms to achieve high‑precision space‑time synchronization and enable mmWave/THz ultra‑spot detection, which has not been explored previously. Building on this foundation, we develop a new algorithm capable of identifying ultra‑spots located near the UAV's flight trajectory while simultaneously detecting pose and directional deviations to support adaptive flight correction. Furthermore, our work uniquely integrates RSSI measurements with phase‑variation analysis to significantly enhance the accuracy of ultra‑spot localization. Finally, we leverage the spatial diversity inherent in the multi‑Wi‑Wi architecture to improve ultra‑spot detection reliability and mitigate packet losses caused by UAV body shadowing, an issue often overlooked in prior studies. Together, these innovations establish a comprehensive and fundamentally new approach to high‑accuracy sensing in UAV‑borne ultra‑spot detection systems.}

The rest of this paper is organized as follows: Section II formulates the research problems by presenting and addressing issues related to the detection of mobile ultra-spots. Section III presents the proposed methods to address the issues discussed in Section II. Section IV details the UAV-to-UAV experiment setup and presents the experimental results. Finally, Section V summarizes the research, outlining the contributions and discussing the implications for mmWave UAV-to-UAV communication.

\section{\textcolor{black}{System Model}}
\label{problemformulation}
Figure\,\ref{fig:01} illustrates the concept in which in-flight UAV-B detects the location of the high-speed 60\,GHz communication area (ultra-spot) created by UAV-A from a long distance. To perform high-speed communication during flight, UAV-B needs to fly past UAV-A at a distance that is neither too close to ensure safety nor too far to stay within the mmWave communication range, and at a correct pose that guarantees beam alignment between the transmitter (TX) and receiver (RX). Therefore, once the location of the ultra-spot is identified, the flight direction and trajectory of UAV-B must be adjusted to approach the ultra-spot of UAV-A. Both UAV-A and UAV-B are currently airborne. UAV-B is flying, while UAV-A is either hovering or may also be flying, depending on the flight scenario. In addition, since we conducted the UAV communication experiments in a real-world environment, the experimental data reflect position and altitude errors caused by UAV oscillations due to wind. However, \textcolor{black}{position and velocity errors due to wind effects, as well as sensor errors, were also neglected to simplify the system model}.

\subsection{\textcolor{black}{System Model (1): Long-distance pose correction via relative pose deviation detection}}

\textbf{Scenario:} UAV-A is hovering, while UAV-B is flying in the vicinity, within a range of 50\,m to 200\,m. UAV-B follows a predefined flight path, which can be flexibly adjusted upon detecting any relative pose deviation with respect to UAV-A.

\textbf{Objective:}
Formulate an optimization or estimation problem that allows UAV-B to minimize the deviation between its intended relative pose and the actual relative pose with respect to the ultra-spot orientation created by UAV-A.

Let:
\begin{itemize}
    \item $\mathbf{p}_A \in \mathbb{R}^3$: 3D position of UAV-A (assumed constant since it's hovering)
    \item $\mathbf{p}_B(t) \in \mathbb{R}^3$: 3D position of UAV-B at time $t$
    \item $\mathbf{R}_A, \mathbf{R}_B(t) \in SO(3)$: orientation (rotation matrices) of UAV-A and UAV-B, respectively. $SO(3)$ is the Special Orthogonal group in 3 dimensions, defined as:
    \begin{equation*}
    SO(3) = \left\{ \mathbf{R}\in \mathbb{R}^{3 \times 3}\ \Big|\ \mathbf{R}^\top \mathbf{R} = \mathbf{I},\ \det(\mathbf{R})=+1 \right\}
    \end{equation*}
    \item $\mathbf{T}^{\text{desired}}_{AB}$: desired relative transformation (pose) from A to B
    \item $\mathbf{T}_{AB}(t) = [\mathbf{R}_{AB}(t), \mathbf{p}_{AB}(t)]$: actual relative pose from A to B at time $t$, where $\mathbf{p}_{AB}(t) = \mathbf{p}_B(t) - \mathbf{p}_A$
\end{itemize}

Then, define the relative pose deviation cost:

\textcolor{black}{ 
\begin{equation}
\begin{aligned}
J_1(t) &= \left\| \mathbf{p}_{AB}(t) - \mathbf{p}_{AB}^{\text{desired}} \right\|^2 \\
       &\quad + \lambda \cdot \text{AngleError}\!\left(\mathbf{R}_{AB}(t), \mathbf{R}_{AB}^{\text{desired}}\right)
\end{aligned}
\end{equation}}

\textbf{Goal}: Find control inputs $\mathbf{u}_B(t)$ to UAV-B such that:
\begin{equation}
\min_{\mathbf{u}_B(t)} J_1(t), \quad \text{subject to UAV-B dynamics and constraints}
\end{equation} 

\textbf{Assumptions}: UAV-A is assumed stationary, serving as a localization anchor. The relative pose (translation and orientation) is observable via onboard sensors or communication. Pose deviations are assumed detectable with sufficient precision within 50–200\,m. Since vision-based methods struggle to maintain such accuracy at long range, wireless device-based solutions are considered more feasible.

\subsection{\textcolor{black}{System Model (2): Short-distance time and distance estimation for ultra-spot approach}}

\textbf{Scenario:} When UAV-B is close to UAV-A (less than 50\,\text{m}), the objective is to allow UAV-B to approach UAV-A to within a safe separation of 4\,\text{m}, using precise estimation of remaining distance and approach time, ensuring the correct final pose alignment.

\textbf{Objective:}
Formulate a prediction and control problem that estimates the \textit{time-to-approach} and \textit{distance-to-approach}, ensuring alignment with the desired pose and collision avoidance through safe-distance maintenance.

Let:
\begin{itemize}
    \item $d(t) = \| \mathbf{p}_B(t) - \mathbf{p}_A \|$: Euclidean separation of the two UAVs at time $t$
    \item $v_B(t)$: instantaneous velocity of UAV-B toward UAV-A
    \item $\hat{T}_{\text{approach}}$: estimated time until $d(t) = d_{\text{safe}} \approx 4$\,\text{m}
    \item $\mathbf{T}_{AB}^{\text{target}} = [\mathbf{R}_{AB}^{\text{target}}, \mathbf{p}_{AB}^{\text{target}}]$: desired final pose (at 4\,\text{m} from UAV-A)
\end{itemize}

Then, define the approach time estimation and final pose error:
\begin{equation}
\hat{T}_{\text{approach}}(t) = \frac{d(t) - d_{\text{safe}}}{v_B(t)}
\end{equation} 

\textcolor{black}{
\begin{equation}
\begin{aligned}
J_2 &= \left| d(t_{\text{stop}}) - d_{\text{safe}} \right|^2 \\
    &\quad + \mu \cdot \text{AngleError}\!\left(
    \mathbf{R}_{AB}(t_{\text{stop}}), \mathbf{R}_{AB}^{\text{target}}
    \right)
\end{aligned}
\end{equation}}

\textbf{Goal}:
Design a control strategy $\mathbf{u}_B(t)$ such that:

\begin{align*}
    &\lim_{t \to t_{\text{stop}}} d(t) \to d_{\text{safe}} \\
    &\mathbf{T}_{AB}(t_{\text{stop}}) \to \mathbf{T}_{AB}^{\text{target}} \\
    &J_2 \to \min
\end{align*}

\textbf{Constraints}: 
Safe operation requires \( d(t) \geq d_{\text{safe}} \geq 4\,\text{m} \), 
control inputs within physical limits, and avoidance of overshoot or oscillation that may affect pose alignment or safety.  

\textbf{Assumptions}: First, precise short-range sensors such as LiDAR, stereo vision, and UWB are assumed to be available for estimating relative pose. However, these sensors are often expensive and require substantial processing capabilities onboard the UAV system. Second, real-time control of UAV-B is assumed to be feasible with bounded latency, thereby enabling timely and responsive trajectory adjustments. Finally, the position of UAV-A is supposed to be known with high certainty, allowing it to serve as a reliable reference point for correcting UAV-B’s motion.

\section{{Proposed method}}
\subsection{Space-time synchronization in ultra-spot tracking and its realization using Wi-Wi devices on UAVs.} 

Due to the ultra-narrow beam width nature of mmWave/THz communications, beam tracking is highly sensitive to any change in the \textcolor{black}{position or orientation of the mobile device or antenna \cite{sanchez2020millimeter,xiao2021survey}}. For instance, with a mmWave beam width of 1.4\,m, the opportunity for a UAV traveling at 20\,km/h to pass through and communicate using the mmWave ultra-spot is 397\,ms, not accounting for positional errors due to sensor errors, wind, and antenna oscillations. This requires that the UAV’s flight control and beam steering must be synchronized in both time and distance with the GSs or other UAVs, because any deviation or delay in timing could result in communication failure.

\textcolor{black}{GS is originally an abbreviation for Ground Station, and the term GSs indicates multiple fixed ground stations. This concept is used to distinguish them from UAVs, which are aerial and mobile devices. While GSs generate static communication ultra-spots with predetermined positions and communication angles, UAVs, due to their high mobility, create dynamic ultra-spots whose positions and angles continuously change over time. This makes the detection of ultra-spots generated by UAVs significantly more challenging compared to detecting the fixed ultra-spots created by ground stations \cite{li2024ground}}.

Regarding spatial synchronization, it is not just about measuring distance, but also about maintaining a constant distance between the two entities communicating \textcolor{black}{\cite{isogai2023extremely}}. For example, at time $t$, if two UAVs, A and B, are 5\,m apart in inter-UAV distance and are in a communication hovering state with an antenna angle of 45$^\circ$, spatial synchronization ensures that whenever either UAV shifts its 3D position, including altitude, the other UAV immediately compensates so their inter-UAV distance remains constant, keeping the communication link aligned and uninterrupted. 

As mentioned in Section I, Wi-Wi technology enables simultaneous time synchronization of multiple devices with nanosecond-level accuracy, as well as spatial synchronization (distance measurement) with millimeter-level precision. \textcolor{black}{In this investigation}, we present the concept of using phase variation values and RSSI of multi-Wi-Wi, which are continuously measured and recorded by the four Wi-Wi devices installed on UAV-B, to infer position and estimate the timing for approaching the ultra-spots. By analyzing the RSSI from multiple devices, it is possible to diagnose deviations in the flight direction. Moreover, by measuring the phase variations of multi-Wi-Wi during flight, the inter-UAV distance from the UAV to the ultra-spots can be estimated. The detailed calculation methods are presented in the following subsections.

\subsection{Ultra-spot proximity detection and pose adjustment using RSSI footprints in a dual-UAV Wi-Wi System}
\label{secIII_B}
\begin{figure}[!t]
	\centerline{\includegraphics[width=0.8\columnwidth]{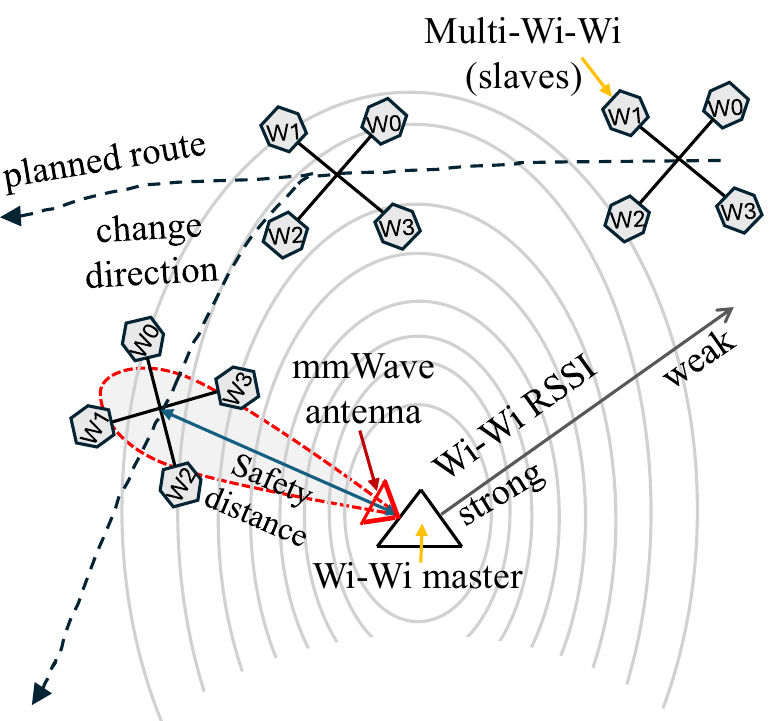}}
	\caption{The concept of ultra-spot proximity detection and pose correction is based on continuously measuring the RSSI of the four Wi-Wi slave devices mounted on UAV-B and comparing these measurements with a predefined RSSI pattern determined by the orientation of the mmWave antenna and the Wi-Wi master mounted on UAV-A.}
	\label{fig:02}
\end{figure}

\textcolor{black}{This research} investigates the use of RSSI measurements collected from four Wi-Wi devices installed on a UAV to enable the UAV to estimate the direction to the ultra-spot while still at a considerable distance from the target and to detect any deviation promptly for timely trajectory adjustments. Fig.\,\ref{fig:02} illustrates the concept of correcting the pose of UAV-B by comparing the RSSI strength of the four Wi-Wi slave devices with that of the Wi-Wi master device mounted on UAV-A. The measured RSSIs are matched against pre-surveyed templates to determine whether UAV-B is approaching the ultra-spot as planned. When a deviation is detected, the flight path is adjusted. As UAV-B nears the ultra-spot, the proposed algorithm (detailed in a later section) calculates the precise timing and distance at which UAV-B will enter and exit the ultra-spot, enabling high-speed communication.
 
In a dual-UAV system, UAV-A remains hovering and is equipped with a single Wi-Wi wireless device operating at 920\,MHz, acting as the master, while UAV-B is flying past UAV-A at a lateral trajectory with a predefined safety distance of $4\,\text{m} \lesssim a \lesssim 5\,\text{m}$. UAV-B is equipped with four Wi-Wi slave devices, mounted at four distinct positions to ensure spatial diversity, thereby allowing for richer and more directional signal interaction with UAV-A.

Each slave Wi-Wi device receives wireless signals from the UAV-A master and continuously reports the RSSI. At any given time, the four RSSI values collectively form a 4-element RSSI signature representing the measured signal strength footprint:

\begin{equation}
\mathbf{r}(t) = [r_1(t), r_2(t), r_3(t), r_4(t)]
\end{equation}

\subsubsection{Proximity estimation based on RSSI magnitude}
As RSSI decreases with increasing distance due to path loss, shadowing, and environmental interference, UAV-B can use the aggregate magnitude of the RSSI values to determine whether it is within a close range of UAV-A, typically under 25\,m. For instance, the weighted sum or average of the four RSSI values can be compared against a predefined threshold:

\begin{equation}
\frac{1}{4} \sum_{i=1}^{4} r_i(t) > \text{RSSI}_{\text{threshold}} \Rightarrow \text{``Close to UAV-A''}
\end{equation}

If the above condition is satisfied, UAV-B is determined to have approached UAV-A at a relatively close distance.

\subsubsection{Pose verification and correction based on RSSI footprint matching}
Once proximity is confirmed, the next step is for UAV-B to adjust its pose to approach UAV-A with the correct alignment of the TX/RX antenna pair mounted on UAV-A and UAV-B in preparation for communication. UAV-B compares its current RSSI footprint $\mathbf{r}(t)$ to a pre-calibrated reference pattern $\mathbf{r}_{\text{ref}}$ \textcolor{black}{\cite{mostafa2025survey}}, which corresponds to the correct pose (orientation) and position required for effective data communication with UAV-A. These reference footprints are constructed during training or calibration flights and represent the ``ideal'' spatial RSSI configuration when UAV-B is correctly aligned with UAV-A.

\textcolor{black}{UAV-B compares its current RSSI footprint with the pre-calibrated reference pattern $r_{\mathrm{ref}}$, where $r_{\mathrm{ref}}$ is obtained by positioning UAV-B at a safe distance from UAV-A, approximately $4.24\,\mathrm{m}$, with the mmWave antenna angles preset so that the beam alignment is precisely calibrated. At this position, four RSSI values are recorded from the four Wi-Wi slave devices on UAV-B (receiving signals transmitted from the Wi-Wi master on UAV-A). Since the RSSI value at each Wi-Wi device differs at any given moment, when UAV-B is in the correct pose, a characteristic pattern is observed, for example, Wi-Wi~03 receives the strongest RSSI from the master, followed by Wi-Wi~01, then Wi-Wi~02, and so on. These four RSSI values, after being ranked from strongest to weakest, form the reference pattern. During experimental flight, UAV-B simply detects any deviation in the RSSI strength ranking among the four Wi-Wi devices and immediately adjusts its heading to the direction where the RSSI distribution across the four devices most closely matches the reference pattern $r_{\mathrm{ref}}$}.

If the ordering of RSSI values deviates from the expected order in the reference pattern, (e.g., $r_3 > r_2 > r_0 > r_1$ as an example in Fig.\,\ref{fig:02}), it is inferred that UAV-B’s orientation or position is suboptimal. The mismatch suggests that the antennas are not aligned as expected, which may lead to degraded data reception or communication failure.

UAV-B then estimates the directional gradient of the deviation (i.e., which RSSI values are stronger/weaker than expected) to infer the approximate direction of UAV-A, and executes corrective maneuvers in its flight path and pose:

\begin{itemize}
    \item If RSSI pattern $\neq$ reference: estimate correction direction and adjust flight trajectory and attitude.
    \item If RSSI pattern $\approx$ reference: maintain pose.
\end{itemize}

\begin{align*}
\text{If } \operatorname{order}(\mathbf{r}(t)) &\neq \operatorname{order}(\mathbf{r}_{\text{ref}}) \\
&\Rightarrow \text{``Adjust pose or approach direction''}
\end{align*}

\subsubsection{Exit detection based on RSSI drop}

When UAV-B starts moving away from UAV-A, all four RSSI values will decrease consistently. Once the average or minimum RSSI value falls below a specified lower threshold, it is inferred that UAV-B has exited the vicinity of UAV-A:

\begin{align*}
\min_i r_i(t) < \text{RSSI}_{\text{exit}} \Rightarrow \text{``Outside proximity''}
\end{align*}

However, in practice, the RSSI signal fluctuates considerably and is not fully reliable \textcolor{black}{\cite{yang2024positioning}}; therefore, it is used only for pose correction and for estimating when UAV-B is approaching UAV-A from a relatively long distance (e.g., beyond 25\,m). For distances below 25\,m, to accurately determine the distance and the timing of entering and exiting the ultra-spot, we use the phase variation values of Wi-Wi instead of RSSI. Details are provided in the following section.

\subsection{Estimate distance to ultra-spot approach from in-flight UAV, by exploiting Wi-Wi's phase variation values}

\begin{table}[!t]
\caption{Notation and functions used throughout the paper}
\centering
\resizebox{\columnwidth}{!}{
\begin{tabular}{|c|c|}
\hline
 \textbf{Symbol} & \textbf{Definition} \\ \hline
$d_0$ &  \begin{tabular}{@{}c@{}} Inter-UAV distance between UAV-A and UAV-B\\ at the starting position.\end{tabular} \\ \hline
$\Delta_{dt}$  & Distance variation calculated from Wi-Wi's phase variation.\\ \hline
$d_t$ & \begin{tabular}{@{}c@{}}Inter-UAV distance between UAV-A and UAV-B\\ when UAV-B closely approaches the ultra-spot.\end{tabular}   \\ \hline
$d_s$ & \begin{tabular}{@{}c@{}}Inter-UAV distance between UAV-A and UAV-B\\ when UAV-B starts to update the speed.\end{tabular}   \\ \hline
$S$ & Center position of ultra-spot of UAV-A  \\ \hline
$w$ & Half width of ultra-spot \\ \hline
$a$ & Safety distance \\ \hline
$h$ & \begin{tabular}{@{}c@{}}Altitude difference between UAV-A, UAV-B\\ at the center of the spot.\end{tabular} \\ \hline
$p_{\mathrm{sc}}$ & The distance from UAV-B to the ultra-spot center.\\ \hline 
$p_t$  & \begin{tabular}{@{}c@{}}The distance that UAV-B has traveled \\ from the starting point.\end{tabular} \\ \hline
$p_s$  & The distance used to calculate the updated speed.\\ \hline
$v_s$ & Updated speed used to estimate ultra-spot entry/exit time.\\ \hline
$s$ & Time when UAV-B gets Wi-Wi data for reference speed\\ \hline
$R_0$, $R_s$, $R_t$ & \begin{tabular}{@{}c@{}}Positions of UAV-B at the start, at reference speed,\\ and at the time when the estimator is initiated.\end{tabular} 
\\ \hline
$t_{\mathrm{ent}}, t_{\mathrm{exit}}$ & Estimated times until ultra-spot entry/exit.\\ \hline
\end{tabular}
}
\label{tab1}
\end{table}

\label{distospot_es}
 \begin{figure}[!t]
	\centerline{\includegraphics[width=1\columnwidth]{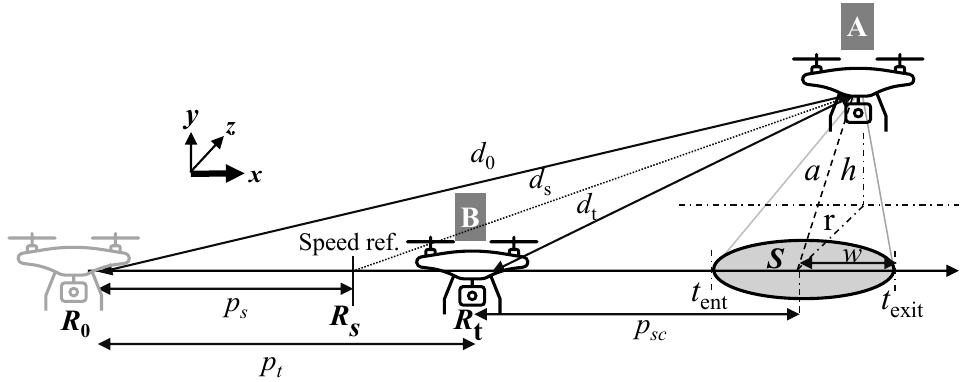}}
	\caption{Two scenarios estimate the time UAV-B flies into the mmWave ultra-spot of UAV-A. ($i$) Scenario 1: When UAV-B flies at a velocity of $v_s$ along the $x$-axis before entering the ultra-spot, while UAV-A anchors at a fixed position. ($ii$) Scenario 2: When UAV-A and UAV-B both fly at a velocity of $v_s$ in opposite directions before entering the ultra-spot.}
	\label{fig:03}
\end{figure}

In principle, a Wi-Wi slave device can use its phase variation data to estimate the distance it has moved relative to the Wi-Wi master device. If the master and slave are placed on two UAVs, intending to perform high-speed communication at the ultra-spot, we can determine whether a UAV is approaching or moving away from the ultra-spot location. However, in practice, relying on data from a single Wi-Wi device is unreliable because packet drops can cause significant errors in the phase variation values. Consequently, using multi-Wi-Wi helps determine the distance and direction to the ultra-spot more reliably and leverages the diversity in calculation results. 

 We proposed two ultra-spot detection algorithms for two flight scenarios, described in Fig.\,\ref{fig:03}. All the parameters illustrated in Fig.\,\ref{fig:03} and used in the proposed algorithm are summarized in Table\,\ref{tab1}. In both scenarios, UAV-B triggered the time estimator from 2.2 $\sim$ 3.7$\,$\,s before approaching the ultra-spot area. Before starting the flight, we used the RTK-GNSS 3D positions (latitude, longitude, and altitude) of UAV-A and UAV-B, shared in the network by drone mapper technology \cite{nguyen2025prediction}, to pre-calculate the initial inter-UAV distance $d_0$ between UAV-A and UAV-B using the Haversine formula \cite{sinnott1984virtues}. Wi-Wi is currently still under research and development, and it requires knowledge of the initial distance between the master and slave devices at the moment of space-time locking to estimate their inter-UAV distances during movement. However, this procedure only needs to be executed once before both UAVs start flying. After that, it is sufficient to rely solely on the Wi-Wi phase variation data to determine the inter-UAV distance between the two UAVs during flight, provided that the flight direction has been fixed, for example, flying along the $x$-axis. In this way, we use it to estimate the distance at which UAV-B approaches and exits the ultra-spot, also known as ultra-spot detection.
 
The relative distance $d$ between two Wi-Wi devices is proportional to the propagation delay phase $\phi$, expressed as
\begin{equation}
d = -\frac{\lambda}{2} \left( \frac{\phi}{\pi} + K \right),
\end{equation}
where $\lambda \approx 325 \,\text{mm}$ denotes the wavelength of the 
channel employed by Wi-Wi, and $K$ is the integer ambiguity of the phase. 
The observed phase $\phi$ is restricted to the physical phase space 
$\left(-\pi, \pi\right]$. Phase unwrapping is performed with $\pi$ based on phase continuity; for instance, $K$ is incremented when a subsequent phase value crosses the boundary of the phase space.

\begin{equation}
\Delta_{d} =-\frac{\lambda}{2}\left(\frac{\Delta_{\phi}}{\pi}+\Delta_{K}\right)
 \end{equation} 
During the flight, the updated inter-UAV distance between UAV-A and UAV-B is calculated as follows:
\begin{equation}
\label{eq1}
d_t=d_0+\Delta_{\mathrm{d}}
\end{equation}
where $\Delta_{\mathrm{d}}$ represents the distance variation calculated from the phase variation values collected when the Wi-Wi devices move from their original positions. The calculation commences once all Wi-Wi devices have synchronized their clocks and phase tracking values across all devices. Further details regarding the distance variance calculation based on phase values can be found in \cite{isogai2023extremely}.

\textcolor{black}{For scenario 1, in which UAV‑A is hovering and serves as a fixed anchor point,} the distance from UAV-B to the UAV-A's ultra-spot center $p_{\mathrm{sc}}$ is estimated by,
\begin{equation}
\begin{split} 
&a\ =\sqrt{h^2+r^2}\\
&p_{\mathrm{sc}}=\sqrt{d_0^2-a^2}-\sqrt{d_t^2-a^2}
\end{split}
\end{equation}
The updated speed of UAV-B is estimated by
\begin{equation}
v_s=\frac{p_{t\ -}p_s}{t-s}=\frac{\sqrt{d_0^2-a^2}-p_{\mathrm{sc}}-p_s}{t-s}
\end{equation}
The durations that UAV-B enters and exits the ultra-spot are estimated as follows
\begin{equation}
\label{eq:estimatedtime}
t_{\mathrm{ent}}, t_{\mathrm{exit}}=\frac{p_{sc}\pm w}{v_s}
\end{equation}

\textcolor{black}{In scenario 2, predicting a dynamic ultra spot (both UAVs are in motion (cross path flight)),} the parameter $a$ remains constant since both UAVs were at fixed altitudes during the experiment. However, since the flight directions are opposite, the distance between each UAV and the center of the ultra-spot, as well as the updated velocity, are estimated using the following approach:
\begin{equation}
\begin{aligned}
p_{\mathrm{sc}}=\frac{\sqrt{d_t^2-a^2}}{2}\\
v_s=\frac{p_{\mathrm{sc}}(s)-p_{\mathrm{sc}}(t)}{t-s}
\end{aligned}
\end{equation}

Similar to scenario 1, the estimated times of entering and exiting the ultra-spot in scenario 2 are determined using equation (\ref{eq:estimatedtime}). However, scenario 2 leads to reduced communication duration at the common ultra-spot and increased detection errors, since both UAVs contribute to the estimation inaccuracies.

Let the current time be denoted as $t_c$. Let the ground truth time when the UAV enters the ultra-spot be denoted as $t_{\text{entGT}}$, and the ground truth exit time be $t_{\text{exitGT}}$. These values, $t_{\text{entGT}}$ and $t_{\text{exitGT}}$, are determined based on the timestamps of the communication established and communication disconnected events recorded by the mmWave transceivers after the experiment.

For the entry and exit times estimated using Wi-Wi, denote them as $t_{\text{entWi}}$ and $t_{\text{exitWi}}$, where:
\[
t_{\text{entWi}} = t_c + t_{\text{ent}}
\]
\[
t_{\text{exitWi}} = t_c + t_{\text{ext}}
\]

The estimates $t_{\text{ent}}$ and $t_{\text{ext}}$ are derived from the phase variation values of Wi-Wi introduced in Equation~(\ref{eq:estimatedtime}).

Similarly, the ultra-spot entry and exit times estimated using GNSS and RTK-GNSS are denoted as: $t_{\text{entGNSS}}$, $t_{\text{exitGNSS}}$, $t_{\text{entRTK}}$, $t_{\text{exitRTK}}$. These are determined by the method described in the previous work~\cite{nguyen2025prediction}. The error between the estimated time and the actual entry and exit time of the ultra-spot is determined as follows:

\begin{equation}
\label{errors_eval}
\begin{aligned}
\varepsilon_{\text{entWi}} = \left| t_{\text{entWi}} - t_{\text{entGT}} \right|\\
\varepsilon_{\text{exitWi}} = \left| t_{\text{exitWi}} - t_{\text{exitGT}} \right|\\
\varepsilon_{\text{entGNSS}} = \left| t_{\text{entGNSS}} - t_{\text{entGT}} \right|\\
\varepsilon_{\text{exitGNSS}} = \left| t_{\text{exitGNSS}} - t_{\text{exitGT}} \right|\\
\varepsilon_{\text{entRTK}} = \left| t_{\text{entRTK}} - t_{\text{entGT}} \right|\\
\varepsilon_{\text{exitRTK}} = \left| t_{\text{exitRTK}} - t_{\text{exitGT}} \right|
\end{aligned}
\end{equation}

 \begin{figure}[!t]
	\centerline{\includegraphics[width=1\columnwidth]{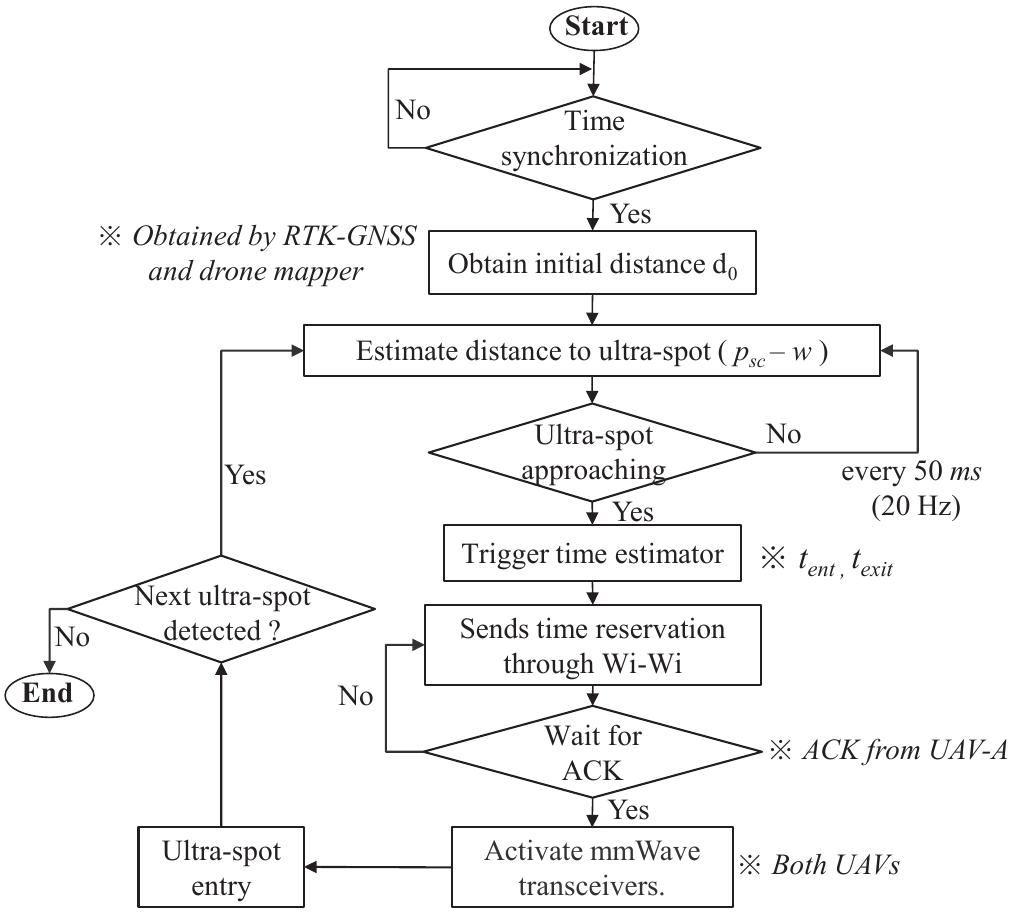}}
	\caption{The algorithm flow chart of the ultra-spot detection process is implemented on UAV-B from the start of the flight until exiting the ultra-spot.} 
	\label{fig:04}
\end{figure}

To control and manage the issue of space-time synchronization among internal control circuit components within a UAV, as well as between multiple UAVs, we use a space-time synchronization unit (STSU), which is a circuit board designed to synchronize the timing of component boards within the onboard UAV system (such as the mmWave transceiver, Wi-Wi, camera, Raspberry Pi, etc.). After the ultra-spot detection algorithm estimates distance and timing for entering and exiting the ultra-spot area, both STSU devices in UAV-A and UAV-B will schedule a future time in the common time axis to activate their mmWave transceivers simultaneously for high-speed communication. The algorithm diagram in Fig.\,\ref{fig:04} illustrates the entire ultra-spot detection program, encompassing time estimation and the utilization of STSU for time scheduling. 

\subsection{Spatial diversity effects of multi-Wi-Wi.}
The shadowing effect occurs due to obstructions from the UAV’s own body during flight, which affect the propagation of wireless signals \cite{khuwaja2018survey,badi2021characterization}, especially in the 920\,MHz band used by Wi-Wi. Theoretically, relying solely on data from a single Wi-Wi module for ultra-spot detection seems sufficient. However, empirical evidence indicates a significant packet drop rate (20\%\,$\sim$\,80\%) in Wi-Wi devices, particularly in the presence of obstacles or environmental interference encountered by the UAV. After packet drops, the phase values transmitted by the Wi-Wi devices become erroneous, consequently impacting ultra-spot detection outcomes. Hence, we suggest collecting data from multi-Wi-Wi devices simultaneously to broaden the data sources. If abnormal values are detected during packet loss and RSSI assessment, data from other Wi-Wi devices exhibiting more reliable metrics will be prioritized for selection. 

\begin{figure}[!t]	\centerline{\includegraphics[width=1\columnwidth]{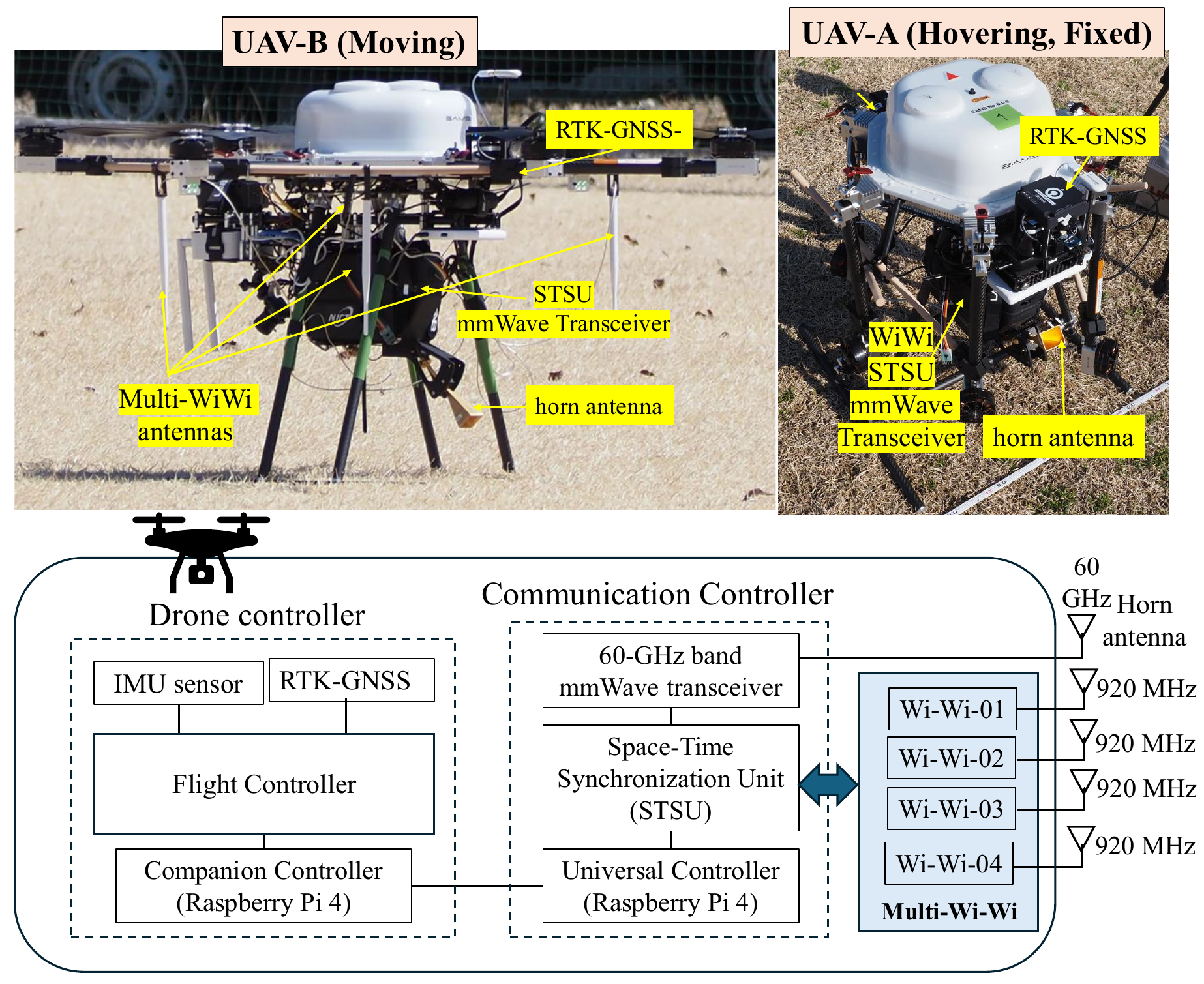}}
	\caption{Image of the hardware setups and schematic of the control system employed in our UAV-to-UAV communication experiments.}
	\label{fig:05}
\end{figure}

\begin{figure}[!t]	\centerline{\includegraphics[width=1\columnwidth]{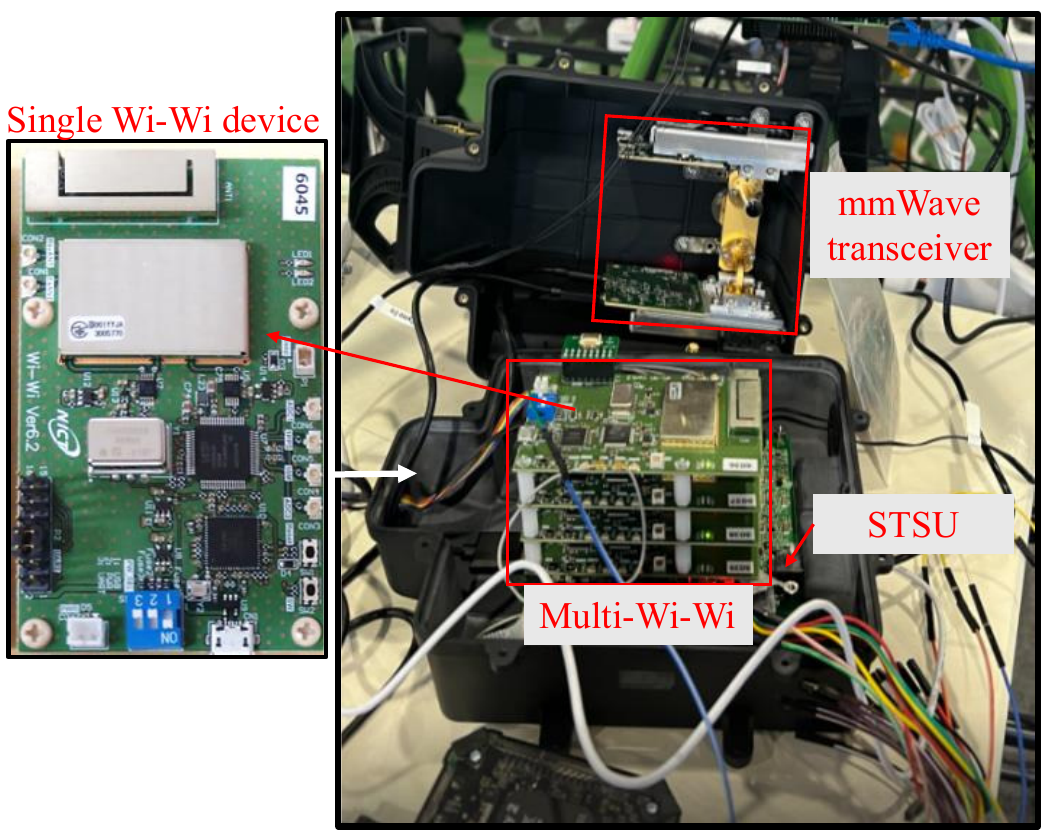}}
	\caption{Internal components and configuration of the hardware box mounted on UAV-B for UAV-to-UAV communication}
	\label{fig:06}
\end{figure}

 \begin{figure}[!t]
	\centerline{\includegraphics[width=1\columnwidth]{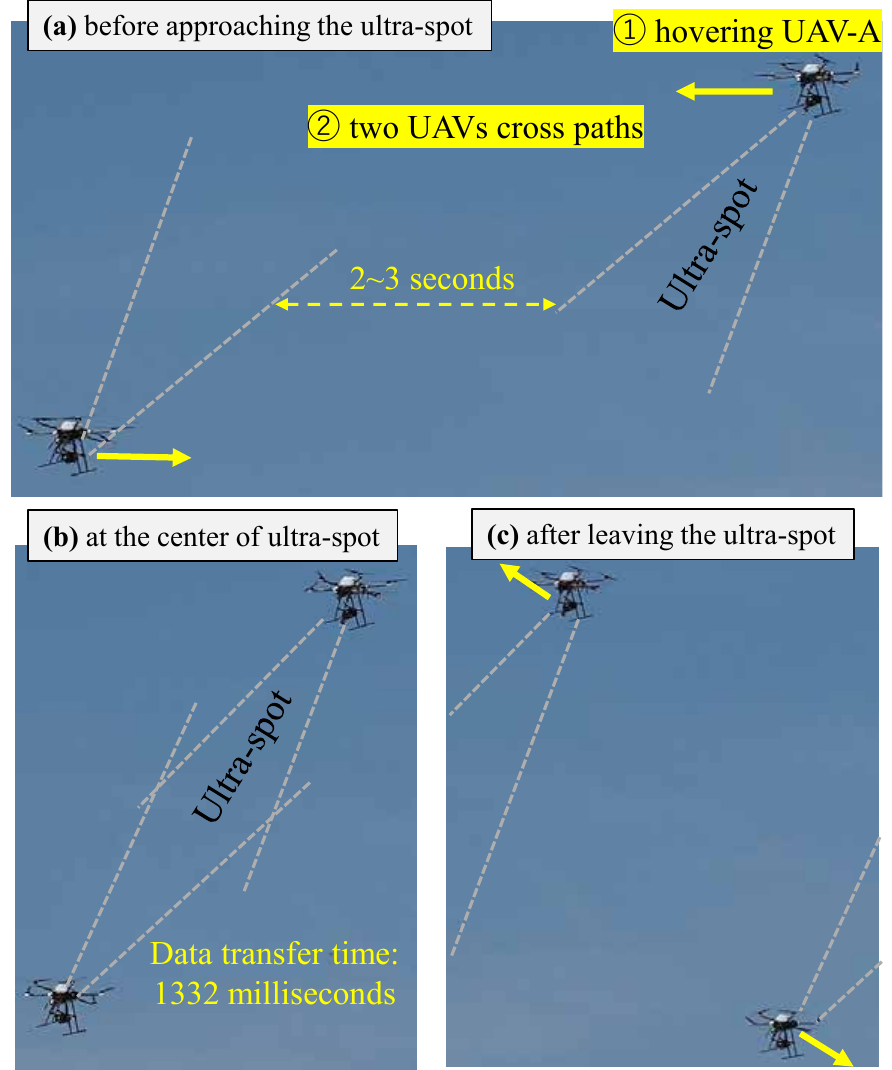}}
	\caption{Experiment of UAVs transmitting large amounts of data at an ultra-spot within a short time period when two UAVs fly past each other, in two scenarios: (1) UAV-A is hovering in the air while only UAV-B flies past the ultra-spot. (2) Both UAV-A and UAV-B are in flight but moving in opposite directions.}
	\label{fig:07}
\end{figure}

\begin{figure}[!t]
    \centering
    \subfloat[]{%
        \includegraphics[width=1\columnwidth]{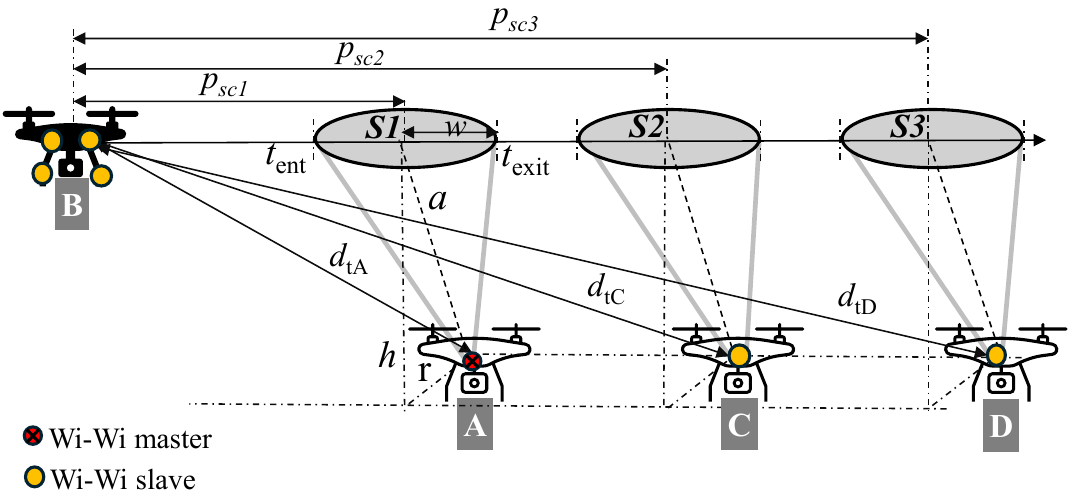}
        \label{fig:8a}
    }\\[1ex]
    \subfloat[]{%
        \includegraphics[width=1\columnwidth]{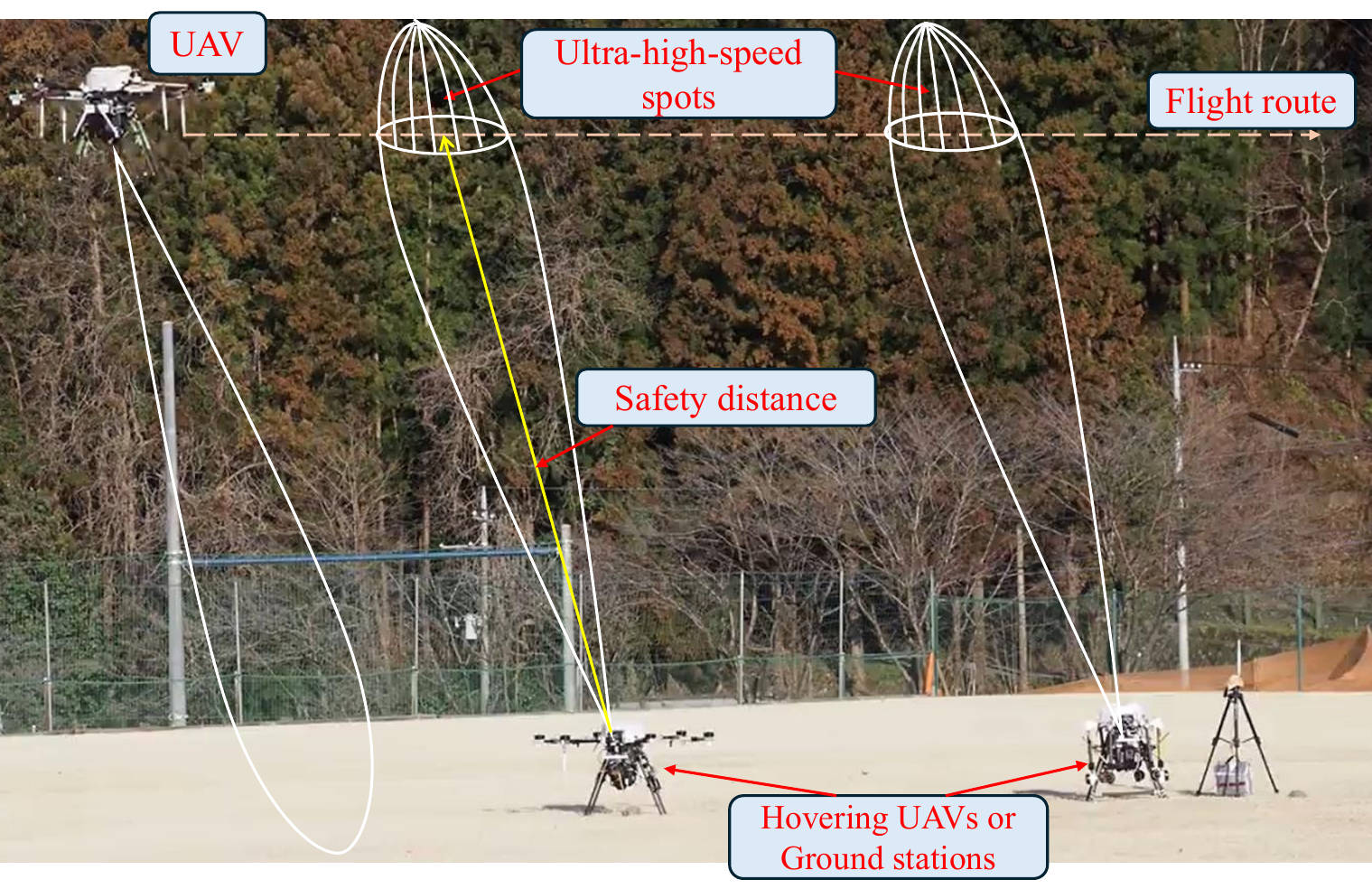}
        \label{fig:8b}
    }
    \caption{(a) Concept of estimating the distance and approaching time to three consecutive ultra-spots from the in-flight UAV-B. (b) The figure illustrates the high-speed communication experiment we conducted between UAV-B in flight and UAV-A, UAV-C, and UAV-D.}
    \label{fig:8}
\end{figure}

The following theoretical analysis proves the spatial diversity benefits achieved by deploying multi-Wi-Wi on a UAV. The basic idea is that using multiple antennas or wireless devices placed at different locations reduces the likelihood of encountering the same interference or obstructions simultaneously. This leads to a decrease in packet loss. The proof is grounded in principles from information theory and communication theory. We can reference the concept of diversity gain \cite{winters1998diversity}, where multiple antennas or devices provide redundancy, thereby improving overall reliability \cite{kildal2004correlation}.

Consider a UAV with multiple antennas or wireless devices positioned at different locations on its body. The UAV's body can cause obstructions, interference, and shadowing, leading to packet loss. We aim to demonstrate that spatial diversity, achieved through the use of multiple antennas, mitigates these adverse effects.

Assume that the UAV has $N$ antennas or wireless devices, each experiencing independent fading and shadowing effects due to obstructions caused by the UAV body \cite{khuwaja2018survey}.

Let $h_i$ represent the channel gain for the 
$i^{th}$ antenna or wireless device. The received signal $y_i$ at the 
$i^{th}$ antenna can be modeled as:
\begin{equation}
y_i=h_ix\ +n_i\ 
\end{equation}
where $x$ is the transmitted signal, $h_i$ is the channel gain for the $i^{th}$ antenna, $n_i$ is the noise at the $i^{th}$ antenna, typically modeled as Gaussian noise with variance $\sigma^2$.

A combining technique, such as maximum ratio combining (MRC), is often employed to utilize spatial diversity \cite{tanbourgi2014effect}. MRC optimally combines the signals from multiple antennas to maximize the Signal-to-Noise Ratio (SNR). The combined signal $y_\mathrm{MRC}$ is given by:

\begin{equation}
y_\mathrm{MRC}=\sum_{i=1}^{N}{w_iy_i}
\end{equation}
where $w_i$ is the weight applied to the $i^{th}$ signal. For MRC, the weights are chosen as: $w_i=\frac{h_i^*}{\left|h_i\right|^2}$,
This ensures that the signals are combined in a way that maximizes the overall SNR ratio.

The combined SNR after MRC can be expressed as:
\begin{equation}
{\rm SNR}_\mathrm{MRC}=\sum_{i=1}^{N}{\left|h_i\right|^2\cdot{\rm SNR}_i}
\end{equation}

where SNR$_i$ is the SNR at the $i^{th}$ antenna. Assuming equal power distribution and independent channels, this becomes:
\begin{equation}
{\rm SNR}_\mathrm{MRC}=\sum_{i=1}^{N}{\left|h_i\right|^2\cdot\frac{P_t}{\sigma^2}}
\end{equation}
where $P_t$ is the transmitted power and $\sigma^2$ is the noise variance.

The probability of packet loss $P_{loss}$ is directly related to the SNR. For a Rayleigh fading channel, the probability of packet loss can be expressed as:
\begin{equation}
P_{\mathrm{loss}}\propto\ \mathrm{exp}\left(-\frac{\mathrm{SNR}}{\gamma}\right)
\end{equation}

where $\gamma$ is a constant depending on the modulation and coding scheme. For $N$ independent channels (antennas), the overall packet loss probability using MRC is:

\begin{equation}
P_{\mathrm{loss, MRC}}\propto\ \mathrm{exp}\left(-\frac{\sum_{i=1}^{N}{\left|h_i\right|^2\cdot{\rm SNR}_i}}{\gamma}\right) 
\end{equation}

Given that $\sum_{i=1}^{N}\left|h_i\right|^2$ represents the diversity gain, the packet loss probability decreases exponentially with the number of antennas.

The UAV body may cause shadowing and block specific wireless devices from receiving a clear signal. However, because each antenna is placed at a different location, it's unlikely all antennas will be obstructed simultaneously. The probability that all antennas experience severe fading or blockage simultaneously is significantly lower than that for a single antenna.

Mathematically, if $P_{obs}$ represents the probability of obstruction for one antenna, the probability that all $N$ antennas are obstructed simultaneously is:

\begin{equation}
P_{\mathrm{obs}, \mathrm{total}}=\left(P_{\mathrm{obs}}\right)^N
\end{equation}

With multiple antennas, $P_{\mathrm{obs},\mathrm{total}}$ becomes significantly smaller, thus improving overall communication reliability.

\subsection{Method for estimating ultra-spot position likelihood using combined RSSI and phase-to-distance values from multi-Wi-Wi on UAV-B}

UAV-B is equipped with four Wi-Wi slave devices, each operating at a carrier frequency of 920 MHz. In contrast, UAV-A is equipped with a single  Wi-Wi device transmitting at the same frequency, and the ultra-spot created by UAV-A serves as the target to be detected. When UAV-B flies across UAV-A, the objective is to evaluate the likelihood that the ultra-spot of UAV-A is located at the position hypothesized by UAV-B. The detection process relies on two types of measurements obtained from the four slave devices mounted on UAV-B: (i) received signal strength indicator (RSSI) values, and (ii) phase measurements, which are further converted into distance estimates. The most reliable detection, corresponding to 100\% certainty, is assumed to be achieved when both RSSI and phase-based distance data from all four slave devices are jointly utilized.

In the generalized likelihood formulation, we define the following hypotheses: 
$H_{1}$, which states that UAV-A is present at the hypothesized location, and $H_{0}$, which states that UAV-A is not at the hypothesized location.

The likelihood ratio is defined as,
\begin{equation}
\label{likelihood01}
\Lambda(\mathbf{X}(t)) = 
\frac{p(\mathbf{X}(t) \mid H_1)}{p(\mathbf{X}(t) \mid H_0)}
\mathop{\gtrless}_{H_0}^{H_1} \eta
\end{equation}

where $\mathbf{X}(t)$ represents the set of observations at time $t$.

\textit{Case 1: RSSI from a single device}: the observed RSSI value over time is $r(t)$.  
\begin{equation}
\label{eqcase1}
\Lambda(r(t)) = \frac{p(r(t) \mid H_1)}{p(r(t) \mid H_0)}
\end{equation}

\vspace{0.5cm}

\textit{Case 2: RSSI from four devices}: the observed RSSI vector over time is $\mathbf{r}(t) = [r_1(t), r_2(t), r_3(t), r_4(t)]$.  
\begin{equation}
\label{eqcase2}
\Lambda(\mathbf{r}(t)) = \prod_{k=1}^{4} \frac{p(r_k(t) \mid H_1)}{p(r_k(t) \mid H_0)}
\end{equation}

\vspace{0.5cm}

\textit{Case 3: Joint RSSI-phase from a single device}: the observed joint RSSI phase value over time is $\mathbf{x}(t) = [r(t), d(t)]^\top$, where $d(t)$ is the distance estimated from phase.  
\begin{equation}
\label{eqcase3}
\Lambda(\mathbf{x}(t)) = \frac{p(\mathbf{x}(t) \mid H_1)}{p(\mathbf{x}(t) \mid H_0)}
\end{equation}

Here, $p(\mathbf{x}(t) \mid H_i)$ is modeled as a \textit{bivariate Gaussian distribution} with mean vector and covariance matrix depending on hypothesis $H_i$.

\textit{Case 4: Joint RSSI-phase from four Wi-Wi devices}: with joint RSSI-phase observation set of 4 Wi-Wi devices given as, 
\[
\mathbf{X}(t) = \{ \mathbf{x}_1(t), \mathbf{x}_2(t), \mathbf{x}_3(t), \mathbf{x}_4(t) \},
\]
where each 
\[
\mathbf{x}_k(t) = [r_k(t), d_k(t)]^\top.
\]

The likelihood ratio for correctly detecting the ultra-spot location is given as:
\begin{equation}
\label{eqcase4}
\Lambda(\mathbf{X}(t)) = \prod_{k=1}^{4} \frac{p(\mathbf{x}_k(t) \mid H_1)}{p(\mathbf{x}_k(t) \mid H_0)}.
\end{equation}

\section{Experimental results}

\subsection{Experiment settings}

Fig.\,\ref{fig:05} illustrates the hardware structure and control system implemented in the experiment. The UAV controller onboard the hexacopters will oversee the operation of the six rotors and manage positioning and recording GNSS and RTK-GNSS data. During the experiment, both UAVs received real-time updates on the positions and altitude of all UAVs through drone mapper \cite{shan2018field,shan2023vehicle}. This enabled us to continuously monitor their statuses, ensuring safety and operational integrity. For high-speed communication, we employed a mmWave transceiver, compliant with the IEEE 802.15.3e standard and operating in the 60\,GHz band for high-speed communication, along with horn antennas for directional signal transmission. Four Wi-Wi antennas were also placed at different positions, underneath the four arms of the 6-rotor type UAV. IMU sensors measure the UAV's orientation and motion parameters. Fig.\,\ref{fig:06} shows the hardware setup inside the hardware box mounted on UAV-B, which is used in our UAV-to-UAV communication experiments. The photo shows the integrated multi-Wi-Wi system, consisting of four single-Wi-Wi devices placed on top of the STSU circuit board. The STSU serves as the central control unit, responsible for precise time synchronization and activation of the mmWave transceiver. 

\textcolor{black}{Currently, the hardware system has been installed on a six-rotor UAV with a wingspan of approximately 1.2\,m, which is capable of carrying a maximum payload of around 5\,kg. Our hardware package including four Wi-Wi slave devices, the STSU, and an mmWave transceiver with antenna, has a total weight of approximately 0.5\,kg. In contrast, the weight of the RTK-GNSS positioning sensor, flight controller, and flight battery depends on the specific UAV platform employed. In general, UAVs with a payload capacity of about 0.5\,kg are sufficient to accommodate the proposed hardware package. Small UAVs can typically be classified into three categories: Mini UAVs, weighing approximately 249–300\,g; Small UAVs, weighing 400\,g–1\,kg; and Medium UAVs (e.g., the Inspire series or small hexacopters), weighing 3–8\,kg. Although mini UAVs are highly optimized in terms of size, weight, and power (SWaP), they are generally unsuitable for mounting a 0.5\,kg hardware module. In contrast, small and medium UAVs provide sufficient structural space for external or under-belly payload integration, and some small UAV platforms include gimbals or mounting structures capable of accommodating compact radar or sensor modules. Therefore, commercially available small and medium UAVs are, in principle, adequate for deploying the proposed hardware system. Finally, regarding power consumption, smaller UAVs are inherently constrained by limited energy capacity, which significantly reduces flight endurance when additional payloads are attached. By comparison, Medium UAVs, such as the Inspire 3 or hexacopter platforms, typically employ more robust battery systems that can better compensate for increased payload weight. Consequently, the proposed solution is considered feasible for practical implementation.}.

Figure\,\ref{fig:07} illustrates an experiment involving high-speed communication between two in-flight UAVs, where the UAVs briefly pass each other. We apply the proposed ultra-spot detection algorithm on UAV-B to estimate the distance to, and approaching/exiting time of, the ultra-spot generated by UAV-A. Two experimental scenarios were tested: one in which UAV-A was hovering and another in which UAV-A was flying. The algorithm was executed on UAV-B approximately \( 2\,\text{s} \leq t \leq 4\,\text{s} \) before it approached the ultra-spot created by UAV-A.

Figure\,\ref{fig:8a} presents the concept of detecting multiple ultra-spots using four Wi-Wi slave devices mounted on the in-flight UAV-B. UAV-A is equipped with a Wi-Wi master device, while UAV-C and UAV-D are each fitted with a Wi-Wi slave device. All the Wi-Wi devices share a common space-time synchronization network. We denote the three consecutive ultra-spots as S1, S2, and S3, and the four Wi-Wi devices on UAV-B as Wi-Wi 01, Wi-Wi 02, Wi-Wi 03, and Wi-Wi 04, respectively. In addition, Fig.\,\ref{fig:8b} shows photographs from our second experiment, which involved UAV-to-ground station (GS) communication. In this setup, we utilized real UAVs to represent the GSs and hovering UAVs to mitigate the impact of wind on position errors, ensuring the stability of mmWave communications. A single UAV-B flies past three GSs (UAV-A, UAV-C, UAV-D), initiating mmWave communication at three ultra-spots created by the GSs. The parameter settings employed in the experiment are as follows: $h$ = 5\,m, $r$ = 5\,m, $w$ = 0.9\,m, $d_0$ = 50.5\,m. We conducted experiments with two UAVs flying at altitudes of 50 m, 75 m, and 100 m, and at speeds of 0.5 m/s, 1 m/s, 2 m/s, and 2.5 m/s. 

Conversely, the second experiment takes place in an anechoic chamber, as illustrated in Fig.\ref{fig:09}. Here, we use six Wi-Wi devices with antennas positioned at six different angles on the UAV arms to verify their efficacy in mitigating obstacles from the UAV bodies and environmental interference. Additionally, we evaluate RSSI loss and packet drops when utilizing different Wi-Wi devices. Fig.\,\ref{fig:09} depicts our investigation into the impact of obstacles from the UAV body on the RSSI and phase variation of Wi-Wi devices, as well as \textcolor{black}{the effectiveness of using data from four Wi-Wi devices simultaneously for ultra-spot detection, as opposed to relying solely on data from a single Wi-Wi device}. In this experiment, one Wi-Wi module functions as the master and is positioned approximately 16.67\,m from the UAV's location, while six Wi-Wi devices placed on the UAV serve as slaves. Following the time and phase synchronization of all Wi-Wi devices, the turntable initiates rotation from $0^{\circ}$ to $360^{\circ}$. Throughout this process, we recorded changes in RSSI and phase variation of each Wi-Wi device as the UAV rotated to simulate the effects of obstacles caused by the UAV's body.

\subsection{Results and discussions}

 \begin{figure}[!t]
\centerline{\includegraphics[width=1\columnwidth]{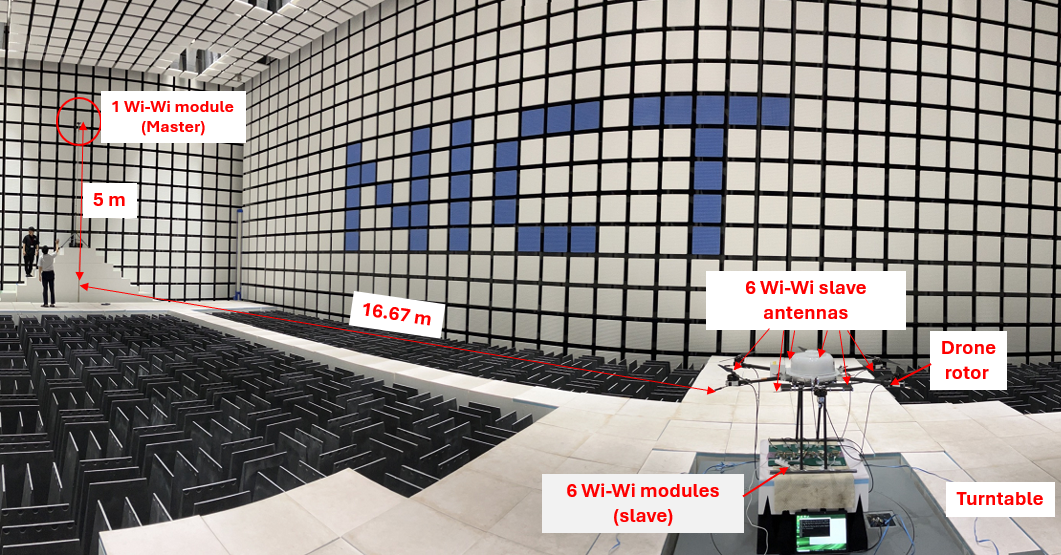}}
	\caption{Photographs of our experiment in an anechoic chamber aimed to examine how obstacles (spatial effects) impact the signal stability of multi-Wi-Wi. The experiment involved surveying the RSSI and phase variation values recorded from six Wi-Wi devices mounted on the UAV's body within an interference-free environment. The UAV was placed on a $360^{\circ}$ rotation table to simulate the effects of obstacles caused by the UAV's body.}
	\label{fig:09}
\end{figure}

 \begin{figure}[!t]
\centerline{\includegraphics[width=1\columnwidth]{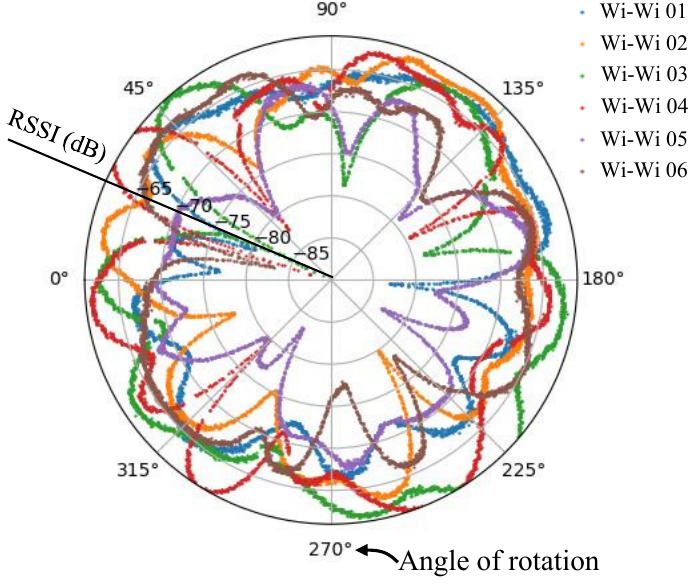}}
	\caption{Impact of UAV body obstructions on RSSI acquired by six Wi-Wi devices when placed on a $360^{\circ}$ rotating table in the anechoic chamber.}
	\label{fig:10}
\end{figure}

 \begin{figure}[!t]
\centerline{\includegraphics[width=0.9\columnwidth]{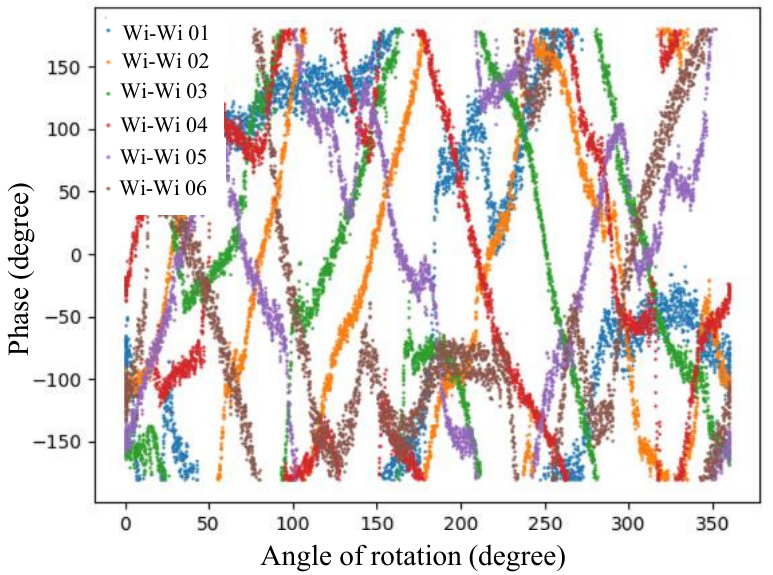}}
	\caption{Impact of UAV body obstructions on phase acquired by six Wi-Wi devices when placed on a $360^{\circ}$ rotating table in the anechoic chamber.}
	\label{fig:11}
\end{figure}

 \begin{figure}[!t]	\centerline{\includegraphics[width=1\columnwidth]{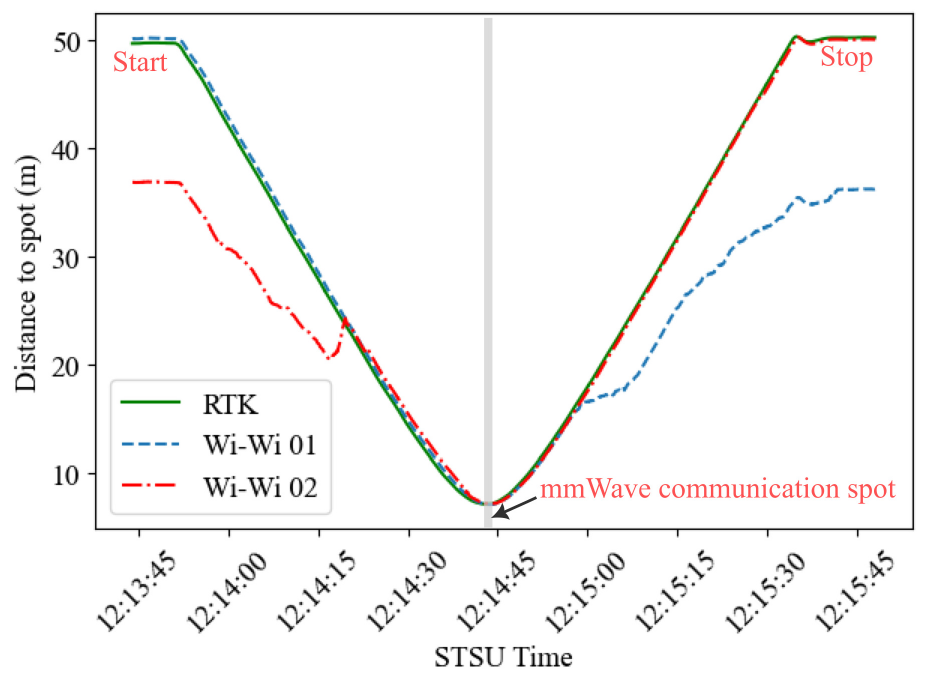}}
	\caption{Compare the distance calculation results utilized for ultra-spot detection between two distinct Wi-Wi devices installed on the UAV. The mmWave ultra-spot is located in the central area of the image, centered around the STSU time stamp of 12:14:45. The ground truth distance, estimated by RTK-GNSS, is provided for reference purposes.} 
	\label{fig:12}
\end{figure}

 \begin{figure}[!t]	\centerline{\includegraphics[width=1\columnwidth]{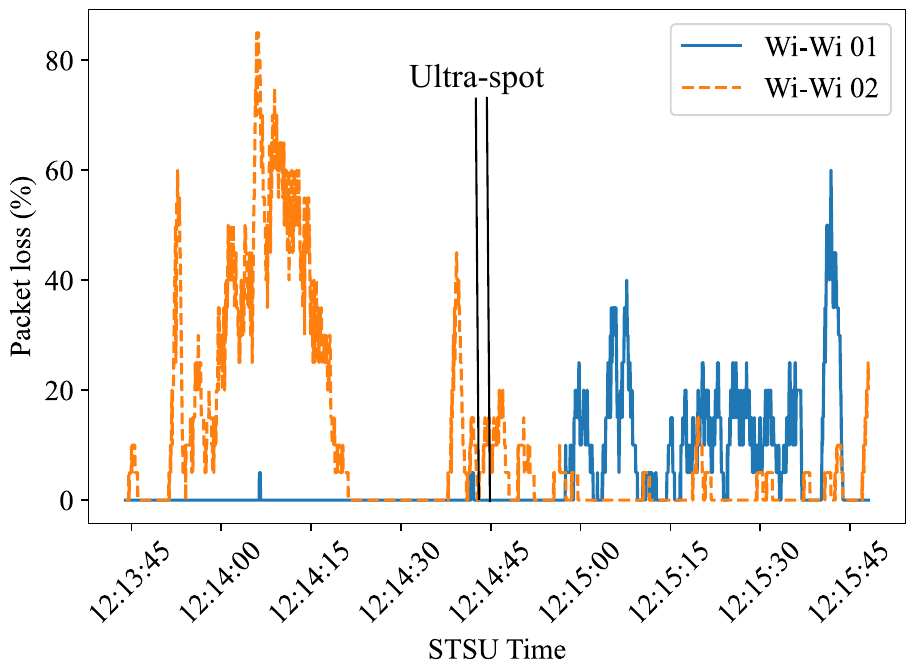}}
	\caption{Packet drop rate of two Wi-Wi devices on UAV-B before entering and after exiting the ultra-spot area. The mmWave ultra-spot corresponds to the central area of the image, around the STSU timestamp 12:14:45.} 
	\label{fig:13}
\end{figure}

% revised done
Figure \ref{fig:10} displays the results from an RSSI survey of six Wi-Wi devices positioned at different locations on the UAV's body, with the UAV on a $360^{\circ}$ rotating table (see Fig.\ref{fig:09}). The experimental results show that despite minimal distance variations between the Wi-Wi master and the six Wi-Wi slaves, none of the devices exhibit a stable RSSI signal. Specifically, between $20^{\circ}$ to $35^{\circ}$ degrees of rotation, the RSSI of Wi-Wi devices $01$, $03$, $04$, and $06$ sharply decreases from -65\,dB to -85\,dB due to obstructions from the UAV's body. In contrast, Wi-Wi\,$02$, and Wi-Wi\,$05$, remain stable under similar conditions. This pattern is also observed at rotation angles of $95^{\circ}$, $135^{\circ}$, $180^{\circ}$, $230^{\circ}$, and $315^{\circ}$, where some Wi-Wi devices consistently experience significant RSSI loss, while others maintain stable RSSI levels.

% revised done
Simultaneously, an analysis of the phase data from six Wi-Wi devices, as shown in Fig.\,\ref{fig:11}, reveals similar obstacle effects on the phase data, consistent with the RSSI survey results. This experimental analysis of RSSI and phase data from the six Wi-Wi devices underscores the importance of using multi-Wi-Wi, which we refer to as the spatial diversity effect in this paper. We applied a method to consistently adjust the phase value within the range of $-180^{\circ}$ to $180^{\circ}$. Notably, as the table rotates $360^{\circ}$, there are instances where the phase value undergoes sudden shifts, resulting in unstable distance calculations. However, stable phase data from a specific Wi-Wi device is consistently available regardless of the rotation angle. Therefore, deploying multi-Wi-Wi in diverse locations is essential for improving the stability of ultra-spot detection.

% revised done
In our UAV-to-UAV communication experiment, we utilized phase data from two Wi-Wi devices to estimate the inter-UAV distances between UAVs for ultra-spot detection. Fig.\,\ref{fig:12} displays the results of assessing the inter-UAV distance from UAV-B to the ultra-spot created by UAV-A, with UAV-A stationary and UAV-B in motion. The flight begins with UAV-B positioned 50\,m away from UAV-A and continues until UAV-B reaches the center of UAV-A's communication ultra-spot, where the distance between the UAVs is 7.07\,m. The flight concludes with UAV-B 50\,m past the ultra-spot, 100\,m from its starting point. Furthermore, Fig.\,\ref{fig:12} presents the ultra-spot detection outcomes obtained with two Wi-Wi devices mounted on UAV-B. The dashed blue line represents the results using only the Wi-Wi\,$01$ device for spot detection. The central area of the ultra-spot, indicated by the concave region in the center of the image, marks the point where the inter-UAV distance is approximately 7.07\,m. The estimation results using the Wi-Wi\,$01$ device align with those obtained by RTK. However, after exiting the ultra-spot, the Wi-Wi\,$01$ device's distance estimation becomes inaccurate compared to RTK due to obstructions from the UAV's body. Conversely, the dashed red line represents the distance estimation using the Wi-Wi\,$02$ device, showing the opposite pattern. Before entering the ultra-spot, the Wi-Wi\,$02$ device's estimates are incorrect compared to RTK, but after exiting the ultra-spot, its estimates match those of RTK. This experimental result demonstrates the effectiveness of using phase values from both Wi-Wi devices to enhance the stability of ultra-spot detection.

% revised done
Figure\,\ref{fig:13} illustrates the packet drop rates of two Wi-Wi devices throughout the flight. It is clear that before entering the designated ultra-spot, Wi-Wi\,02 experiences a notably high packet drop rate, leading to inaccurate ultra-spot detection results. In contrast, Wi-Wi\,01 shows a significantly low packet drop rate before entering the ultra-spot, making its data suitable for ultra-spot detection. However, after exiting the ultra-spot, the packet drop rate of Wi-Wi\,02 drops considerably, while the packet drop rate of Wi-Wi\,01 rises substantially. Therefore, the phase variation data of Wi-Wi\,02 is used to estimate the duration and distance from the UAV to the next ultra-spot. This finding demonstrates the effectiveness of using multi-Wi-Wi to overcome the high packet drop issue inherent in current Wi-Wi technology. We refer to this effect as spatial diversity resulting from using more than one Wi-Wi device on the UAV to improve communication reliability.

\textcolor{black}{In summary, although a single Wi‑Wi device can already provide accurate space–time synchronization and real‑time tracking of inter‑mobility distance variations when combined with high‑precision RTK‑GNSS, as partially discussed in our previous work \cite{nguyen2025prediction}, there are two key reasons for employing four Wi‑Wi devices.
\textit{(i)} Spatial diversity:
as shown in Fig.\,\ref{fig:12} and Fig.\,\ref{fig:13}, a single Wi‑Wi device may experience line‑of‑sight blockage by the UAV body, resulting in packet loss, incorrect phase‑offset estimation, and consequently unreliable space–time synchronization. Placing multiple Wi‑Wi devices at different UAV locations ensures that at least one device maintains an unobstructed propagation path at any given moment, thereby improving synchronization reliability.
\textit{(ii)} Maintaining relative orientation between UAVs:
although at least two Wi‑Wi devices are required, our results show that four are preferable. With only two devices, the relative orientation of two flying UAVs cannot be kept stable, meaning the horn antennas cannot remain aligned throughout the flight. As discussed in Section\,\ref{secIII_B}, using four RSSI footprints enables more robust position identification than using one or two. Furthermore, maintaining four simultaneous inter‑UAV distance measurements (four slave devices paired with one master) provides better control of antenna pointing accuracy by stabilizing the relative attitude of the two UAVs.}

\begin{figure*}[h]
    \centering
    \includegraphics[width=1\textwidth]{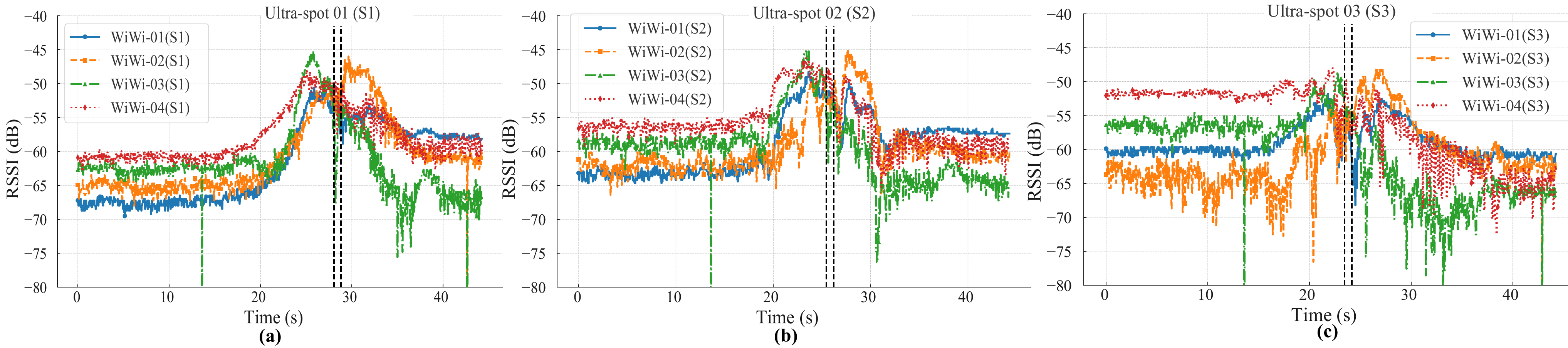}
    \caption{RSSI measurements from four Wi-Wi devices on UAV-B as it flies past three ultra-spots created by three hovering UAVs.}
    \label{fig:14}
\end{figure*}

\begin{figure}[!t]
    \centering
    \includegraphics[width=0.49\textwidth]{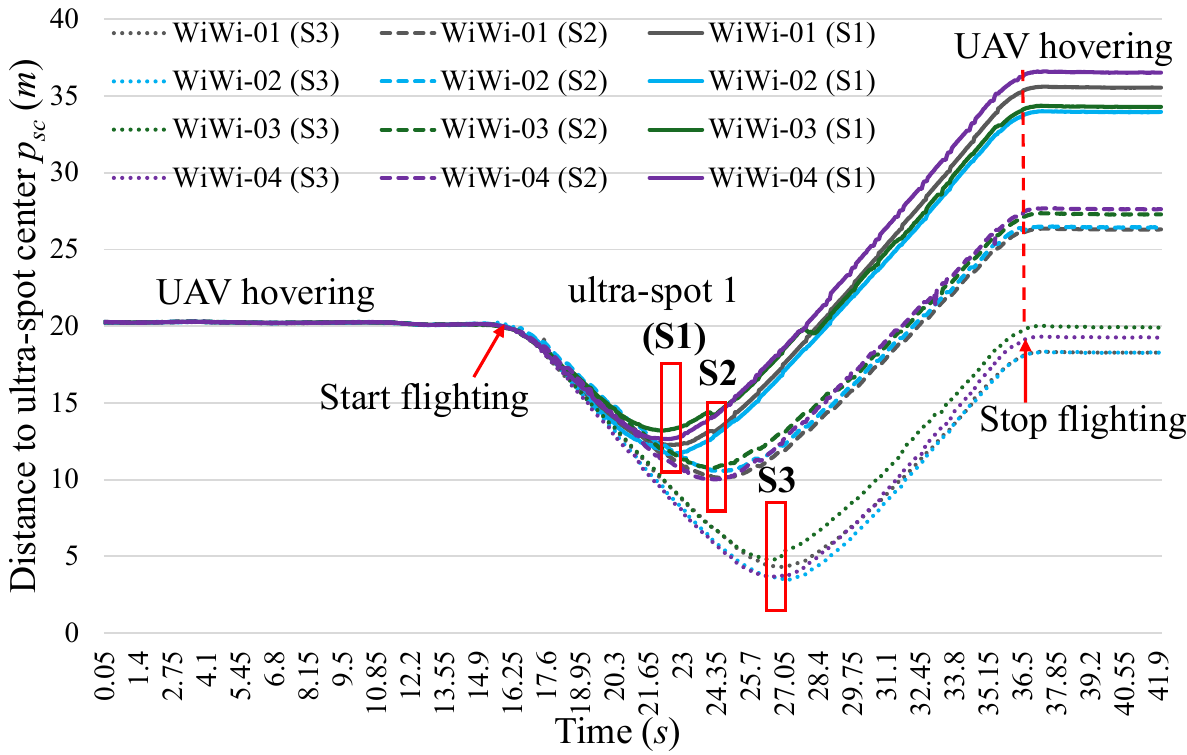}
    \caption{Distances from UAV-B to center of three ultra-spots ($p_{sc}$) created by UAV-A (S1), UAV-C (S2), and UAV-D (S3).}
    \label{fig:15}
\end{figure}

Figure\,\ref{fig:14} shows the RSSI measurements from four Wi-Wi devices placed on the UAV as it flies over three ultra-spots located 3\,m apart. \textcolor{black}{Basically, the RSSI measurement procedure was carried out at a fixed ground station to capture the RSSI variations from the four Wi-Wi devices mounted on the flying UAV-B. The measured RSSI fluctuation throughout the UAV-B’s flight as it passed horizontally over the GS for more than 40 seconds is shown in Figure 14.
Regarding the measurement delay, each Wi-Wi device collects samples at a rate of 20 samples per second, so in principle, the RSSI measurement latency is below 50\,ms. Regarding threshold selection: the RSSI threshold was chosen based on experiments conducted before UAV-B approached the vicinity of the ultra-spot generated by the GS. Since the safety distance to avoid collision is defined as 4.24\,m, we considered the UAV-B to be approaching the beginning of the ultra-spot region when it was approximately 7\,m away in terms of point-to-point relative distance. In Figures 14 and 17, it can be seen that with a flight speed of 1\,m/s, using an RSSI threshold greater than –55\,dB for all four Wi-Wi devices allows us to eliminate areas that are not within the ultra-spot. This is the basis for our threshold selection}. It can be observed that the RSSI strength tends to increase above -55\,dB as the UAV approaches the ultra-spots. Additionally, for ultra-spot '01', the RSSI from Wi-Wi-03 is the most reliable for spot detection. For ultra-spot '02', it is Wi-Wi-04; for ultra-spot '03', it is Wi-Wi-02. This is because the four Wi-Wi antennas of the four Wi-Wi devices are fixed at different positions on the UAV’s arms. Therefore, when the UAV flies over the ultra-spot, depending on its pose at that moment, the obstruction caused by the UAV’s own body and the line-of-sight (LOS) signal path from the UAV-mounted Wi-Wi antenna to the Wi-Wi antenna at the GS of the ultra-spot approach point will determine which Wi-Wi device provides the most reliable RSSI value.

Figure\,\ref{fig:15} is the estimated distance from UAV-B to three ultra-spots created by UAV-A, UAV-C, UAV-D as mentioned in the concept shown in Fig.\ref{fig:02}. However, instead of keeping the three UAVs hovering in the air, where positional shaking due to wind would affect the estimation of the distance to the three ultra-spots, we placed the three UAVs on the ground as three GSs, representing hovering UAVs at fixed positions, and calculated only the distance from the flying UAV-B to these three fixed ultra-spots, as shown in the experimental image presented in Fig.\ref{fig:8}. Phase variation values recorded from 4 Wi-Wi devices on UAV-B. Because 4 Wi-Wi devices are placed in 4 different positions on the UAV in a rectangular shape, each position is 60\,cm apart. Therefore, in the ultra-spot areas, specifically second 23, second 25, and second 27 in Fig.\,\ref{fig:15}, the estimated distance is the average of the estimated distances of the four Wi-Wi devices. This helps increase the stability of distance estimation compared to using only one device. In addition, we confirm that the experimental positioning accuracy of ultra-spots using 4 Wi-Wi devices is equivalent to RTK-GNSS, with an error of less than 30\,cm.

 \begin{figure}[!t]
	\centerline{\includegraphics[width=0.9\columnwidth]{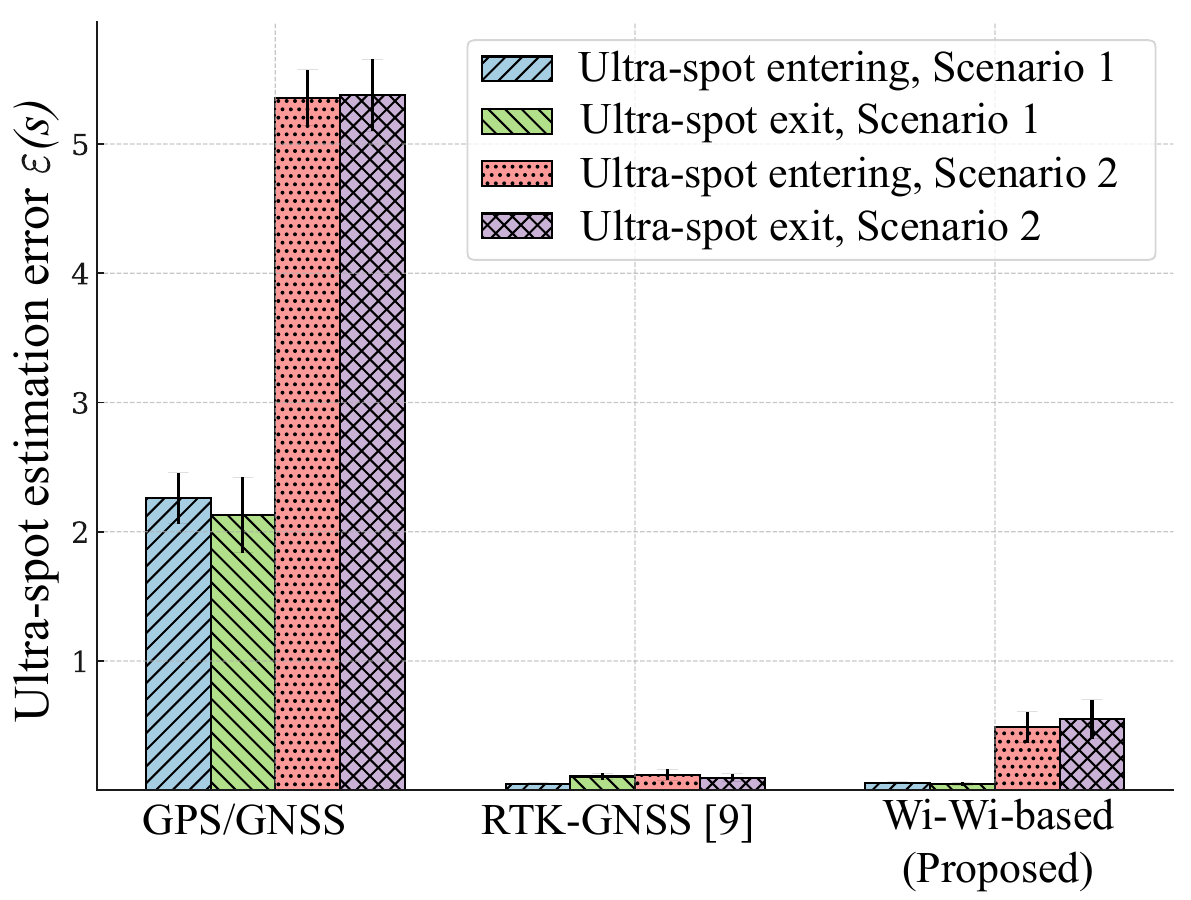}}
	\caption{The error in estimating duration that UAV-B approaches ultra-spot, occurs in two experimental scenarios, mentioned in subsection \ref{distospot_es}.} 
	\label{fig:16}
\end{figure}

 \begin{figure}[!t]	
 \centerline{\includegraphics[width=1\columnwidth]{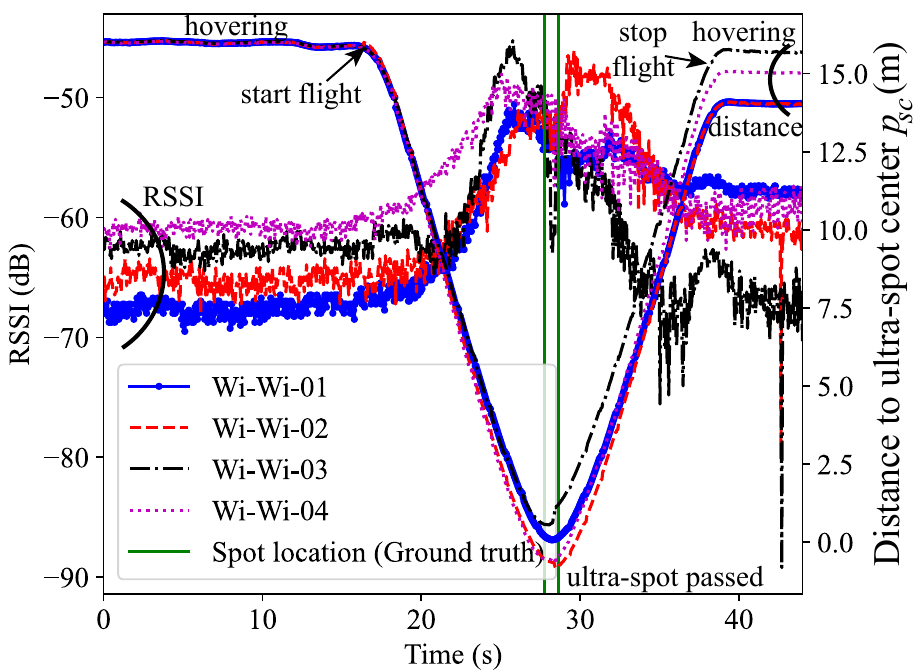}}
	\caption{Joint RSSI-phase measurements of multi-Wi-Wi were conducted on in-flight UAV-B, which started transitioning from hovering state to flying state at second 16.4, approached the ultra-spot at second 27.8, completed the route, and returned to hovering state at second 38.5.}
	\label{fig:17}
\end{figure}

 \begin{figure}[!t]
	\centerline{\includegraphics[width=1\columnwidth]{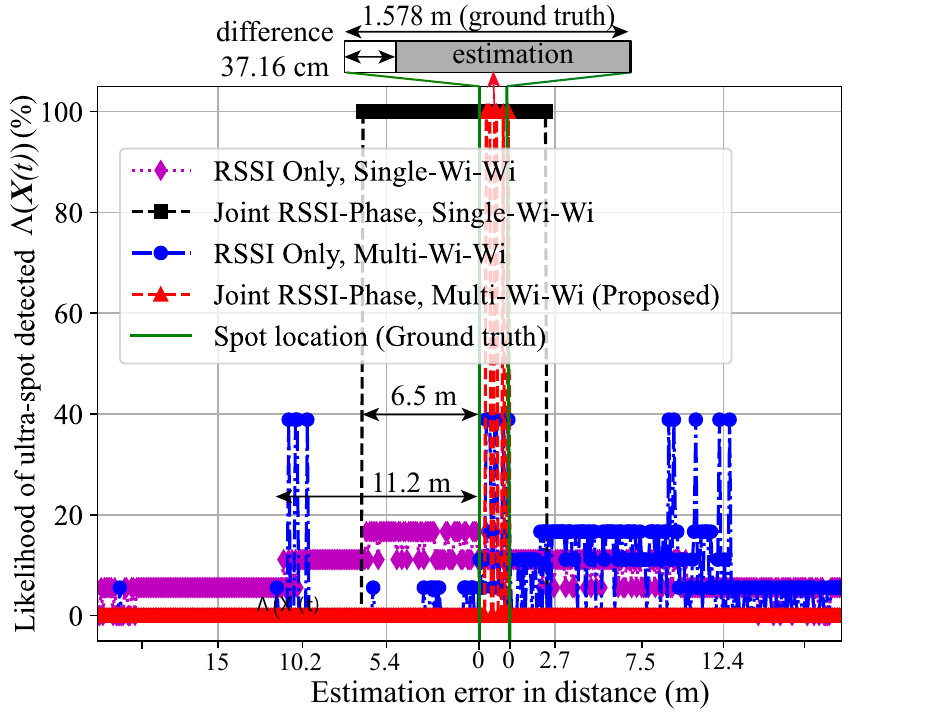}}
	\caption{Detection accuracy of ultra-spots as a function of distance when UAV-B is approximately 20\,\text{m} away from the ultra-spot, and \textcolor{black}{the likelihood that an ultra-spot exists} at the estimated location when applying the proposed algorithm, compared with other approaches and with ground truth.}
	\label{fig:18}
\end{figure}

Figure\,\ref{fig:16} presents a comparison of the estimation errors in the times of entry into and exit from the ultra-spot area using three techniques: GNSS, RTK-GNSS, and multi-Wi-Wi. The estimation error~$\varepsilon$ for each method is calculated using (\ref{errors_eval}), with the six error values corresponding to a single experimental flight scenario. Since two different flight scenarios were conducted, the error evaluation chart in Fig.\,\ref{fig:16} includes a total of 12 error values~$\varepsilon$. We evaluate the results in two flight scenarios. The UAV-B flies at a velocity of approximately 1\,m/s and begins estimating between 2.2\,s, and 3.7\,s before approaching the ultra-spot. The findings reveal that in scenario 1, the ultra-spot detection error for Wi-Wi is 54\,ms, whereas for RTK-GNSS, it is 78\,ms. This indicates that the performance of Wi-Wi is comparable to that of RTK-GNSS and is approximately 40.5 times more accurate than traditional GNSS methods. Conversely, in scenario 2, the average error in estimated ultra-spot exiting time using Wi-Wi is 518\,ms, whereas for RTK-GNSS, it is 108\,ms. Considering an experimental mmWave ultra-spot, which is approximately 1.578\,m wide, the time for a UAV moving at a velocity of 1\,m/s to fly over the ultra-spot is approximately 1.578\,s. The 54\,ms approaching-time error in Scenario 1 (3.42\%) demonstrates the feasibility of the proposed mmWave ultra-spot estimation method. In contrast, the 518\,ms error in Scenario 2 reveals its limitations for counter-directional UAV flights, necessitating algorithmic refinement.

Figure \ref{fig:17} presents the ultra-spot detection results using joint RSSI-phase measurements from four Wi-Wi devices mounted on UAV-B. The actual ultra-spot location (ground truth) does not coincide with the RSSI peak area of any single Wi-Wi device. Conversely, the region with the highest probability of containing the ultra-spot exhibits an RSSI strength ranging from -51\,dB to -56\,dB. This indicates that using only the RSSI from four devices is insufficient to accurately estimate the position of the ultra-spot. Therefore, in addition to relying on RSSI, we also utilized phase variation data, also known as the joint RSSI-phase approach. Regarding phase variation data from the Wi-Wi devices, with initial inter-UAV distance estimated from RTK-GNSS and drone mapper, to assess the distance from in-flight UAV-B to the ultra-spot (also referred to as space synchronization), after compensating for a 4\,m safety margin. It can be seen that the estimated distance to ultra-spot calculated by Wi-Wi-01 is close to 0\,m, corresponding to the average estimated distance of the four Wi-Wi devices, demonstrating that the estimated position of ultra-spot matches the ground truth location with an error of only a few centimeters.

Figure\,\ref{fig:18} illustrates and compares \textcolor{black}{the likelihood in ultra-spot detection of four approaches: (\textit{i}) using RSSI data of a single Wi-Wi device only; (\textit{ii}) using both RSSI and phase data of a single Wi-Wi device; (\textit{iii}) using RSSI data of multi-Wi-Wi; and finally, (\textit{iv}) using RSSI and phase data of multi-Wi-Wi. The confidence index we use is the likelihood ratio at the estimated position where an ultra-spot actually exists, denoted as $\Lambda(\mathbf{X}(t))$. It is computed based on the equations from (\ref{likelihood01}) to (\ref{eqcase4}). We calculated the likelihood ratio for these four approaches based on equations (\ref{eqcase1}), (\ref{eqcase2}), (\ref{eqcase3}), and (\ref{eqcase4}), where each equation corresponds to a specific case. The results show that when using only a single Wi-Wi's RSSI thresholding method, such as a -55\,dB threshold for Wi-Wi-04, the error distance from the actual ultra-spot location is approximately 7.5\,m. Additionally, even with joint phase-RSSI measurement using a single Wi-Wi, the estimated ultra-spot location deviates by 7.09\,m from the ground truth, whereas the actual ultra-spot width in our experiment is 1.578\,m. In contrast, by utilizing only RSSI data, the multi-Wi-Wi approach identifies the three most probable ultra-spot areas with a detection likelihood of 38.9\%, representing a 3.5-fold improvement over a single Wi-Wi approach}. 

\textcolor{black}{%
Figure \ref{fig:18} also presents the estimated ultra‑spot position obtained using the four proposed methods. By computing the distance between the UAV’s position when it first approaches the ultra spot and when communication ends, the proposed joint RSSI–phase multi‑Wi‑Wi method estimates the ultra‑spot width as 1.2064\,m, whereas the ground‑truth width derived from timestamp‑based post‑analysis is 1.578\,m. The method exhibits a 37.16\,cm deviation at the entering position, while the exit position is accurately predicted; this deviation is therefore used as the experimental positioning error, reflecting the limited number of flight trials. Regarding the 186 ms estimation‑time error: the predicted entry position deviates from the ground truth by 37.16\,cm (23.54\% of the true 1.578\,m width), which corresponds to 186\,ms assuming a UAV speed of 2\,m/s. These results demonstrate the feasibility of the proposed method for predicting the ultra‑narrow mmWave communication zone from a flying UAV and indicate its potential for future THz‑band beam‑prediction applications.%
}

\begin{table*}[t]
\caption{\textcolor{black}{Comparison of start-of-the-art mmWave wave positioning, station , mobilities positioning, or related matters.}}
\label{tab:fullpage_mmwave}
\centering
\scriptsize
\setlength{\tabcolsep}{3pt}
\renewcommand{\arraystretch}{1.15}

\begin{adjustbox}{width=\linewidth}
\begin{tabular}{p{2.5cm} p{1.8cm} p{2.8cm} p{2.6cm} p{2.4cm} p{2.6cm} p{3.0cm} p{3.0cm}}
\toprule

\textbf{\textcolor{black}{Method}} &
\textbf{\textcolor{black}{Mobility type}} &
\textbf{\textcolor{black}{Sensing modality}} &
\textbf{\textcolor{black}{Hardware}} &
\textbf{\textcolor{black}{Experimental setup}} &
\textbf{\textcolor{black}{Reported metrics}} &
\textbf{\textcolor{black}{Conditions}} &
\textbf{\textcolor{black}{Key limitations}}\\

\midrule

MiFly \cite{lam20256d} &
UAV in flight &
FMCW radar + IMU fusion &
24\,GHz radar; 200\,MHz BW &
Real drone flights + indoor env. &
Median 3D error 7.2\,cm &
Accuracy degrades at longer range &
Error exceeds 50\,cm beyond ~5\,m \\

\midrule

mmE-Loc \cite{wang2025ultra}&
UAV flight/landing &
mmWave radar + event camera &
COTS event camera + mmWave radar &
30+\,h indoor/outdoor experiments &
Avg localization acc. 0.083 units; 5.15\,ms latency &
Various UAV conditions &
Hardware specs unclear; calibration required \\

\midrule

MultiLoc \cite{blanco2022augmenting} &
Indoor device/user &
CSI-based + FR-based ranging &
60\,GHz mmWave + sub-6 GHz WiFi &
Extensive deployment &
Median localization error 18\,cm &
Reflection ambiguity handling &
Needs sub-6 assistance \\

\midrule

waveSLAM \cite{picazo2023waveslam}&
Mobile robot &
FTM (ToF) + CSI + LiDAR &
60\,GHz radios; URA AoA &
Real prototype deployment &
Errors $<$ 22\,cm; orientation $<$20° &
Optical sensors degrade in mirrors/glass &
Depends on LiDAR/odometry \\

\midrule

mmWave Radio SLAM \cite{rastorgueva2024millimeter}&
Moving UE, fixed BS &
AoA/AoD beam RSSRP &
60\,GHz beam-space model &
Ray tracing + measurements &
RMSE 0.56 ± 0.33\,m &
LOS/NLOS mixed &
Clock-bias estimation issue \\

\midrule

Map-assisted localization \cite{kanhere2019map} &
Moving UE, fixed BS &
AoD + ToF + 3D map &
28\,GHz ray tracing &
Indoor office simulation &
12.6\,cm (LOS), 16.3\,cm (NLOS) &
Explicit LOS/NLOS analysis &
Simulation-driven; needs accurate map \\

\midrule

milliLoc \cite{zhang2023push} &
Target motion &
FMCW radar sensing &
77--81\,GHz FMCW; multi-Rx &
Experiments + simulation &
Median localization; 5.5\,mm ranging accuracy (favorable geometry), degraded at large AoA & 
Sensitive to target AoA and specular reflections; mitigates environmental disturbances using temporal continuity&
Not a full UAV pose estimator; radar-based assumptions differ from communication-beam localization; AoA degrades at large angles.\\

\midrule

SPEBT (vehicular pose + beam tracking) \cite{liu2023successive} &
Vehicle (high mobility) &
Radar odometry + EKF beam tracking &
Navtech CTS350-X radar; ULA &
Simulation on real dataset (Monte Carlo) &
Pose RMSE: (2.77, 4.08)\,m; beam RMSE: AoD 0.22$^\circ$, AoA 4.62$^\circ$ &
AoA harder than AoD; LoS focus &
Large pose drift; not direct UAV positioning \\

\midrule

\textbf{\textcolor{black}{This work}} &
\textcolor{black}{UAV in flight} &
\textcolor{black}{Space-time synchronized (Multi-Wi-Wi, RTK-GNSS assisted)} &
\textcolor{black}{920\,MHz horn antenna, 60\,GHz transceiver} &
\textcolor{black}{Real flights} &
\textcolor{black}{Distance estimation error: $<$ 3\,cm; ultra-spot position error: 37.16\,cm} &
\textcolor{black}{Wind-induced oscillations included; LOS} &
\textcolor{black}{RTK‑GNSS and a drone mapper are needed at the initial stage to determine the distance.}\\

\bottomrule
\end{tabular}
\end{adjustbox}

\end{table*}

\textcolor{black}{Table \ref{tab:fullpage_mmwave} compares the results achieved in this study related to distance estimation and ultra-spot localization, while extending the comparison to other studies on related topics.
It can be seen that, with experimental accuracy in distance estimation using Wi-Wi of less than 3 cm, and under favorable conditions potentially below 1 cm\cite{isogai2023extremely}, this level of accuracy is superior when compared with the positioning and ranging techniques listed in the comparison table.
In addition, regarding the practical accuracy in predicting ultra-spot positions when accounting for variability under real UAV operating conditions (wind, vibration, antenna setup angle errors, hardware limitations, etc.), the value of 0.3716\,m is most directly comparable to papers that report position error in meters under realistic uncertainty sources (multipath effects, outliers, clock bias, motion). Within that subset, the closest published results come from the  mmWave radio SLAM-based approach \cite{rastorgueva2024millimeter}}.

\section{Conclusion}
\textcolor{black}{In this research}, we introduced an mmWave ultra-spot prediction technique based on space-time synchronization using multi-Wi-Wi. The proposed algorithm can identify ultra-spots near the flight path and detect deviations in pose and flight direction by matching the RSSI values of the multi-Wi-Wi system with a pre-surveyed reference pattern. Moreover, the proposed method estimates the distances from mobile UAVs to mmWave ultra-spots and predicts the time required for the UAVs to approach these ultra-spots. We installed multi-Wi-Wi equipped with antennas at various locations on the UAV's body to address the issues of RSSI degradation and unstable phase data caused by obstructions from the UAV itself. To validate the spatial effects of multi-Wi-Wi configurations on ultra-spot detection, two experiments were conducted: one in an anechoic chamber to suppress interference from external radio channels, and another in an outdoor UAV-to-UAV communication setting. The results showed that the UAV can estimate the ultra-spot approaching time with an accuracy of 54\,ms, verifying the feasibility of the proposed method for mmWave ultra-spot prediction. Moreover, by varying the locations and increasing the number of Wi-Wi devices, we significantly reduced packet loss and minimized RSSI and phase signal instability caused by obstacles and interference. This diversification also doubled the stability of ultra-spot detection, both before and after passing through the ultra-spot area. \textcolor{black}{The results presented in this study provide initial evidence of the feasibility of accurately predicting the spatial position of mmWave beams in advance from a flying UAV. This makes it possible for UAVs that store large amounts of sensing data and need to upload such data to ground stations or other UAVs to adaptively adjust their flight direction so that they can approach ultra‑narrow, high‑throughput communication zones with optimal antenna orientation and signal‑reception positioning, thereby enhancing the effectiveness of future mmWave UAV‑to‑UAV communications.}

\section*{Acknowledgement}
The authors gratefully acknowledge Sony Semiconductor Solutions Corporation, our collaborative research partner, for providing the 60\,GHz-band mmWave wireless transceivers. We also wish to thank the members of the Social-ICT System Laboratory and the Space-time Standards Laboratory of NICT for their contributions to the development of the Wi-Wi devices and for their assistance throughout the field experiments. We would also like to thank the Wireless System Laboratory (NICT) for providing the drone mapper equipment used in the UAV experiments. We also wish to express our gratitude to Assoc. Prof. Yoshito Okada (Tohoku University) for supporting the UAV experiments and for his valuable feedback and contributions related to positioning technology based on Wi-Wi.
	
\bibliographystyle{IEEEtran}
\bibliography{IEEEexample}

\end{document}